\documentstyle[pre,aps]{revtex}

\def\be{\begin{equation}} 
\def\ee{\end{equation}}
\def\pa{\partial}
\def\na{\nabla}
\def\lan{\langle}
\def\ran{\rangle}
\def\pr{\prime}
\def\rarrow{\rightarrow}
\def\iff{\infty}

\begin{document}

\title{Smoothing of sandpile surfaces
after intermittent and continuous avalanches:
three models in search of an experiment}

\author{Parthapratim Biswas\footnote{ppb@boson.bose.res.in},
Arnab Majumdar\footnote{Present address: Physics Department, 
Boston University, Boston, MA 02215 USA, E-mail: arnab@bu.edu} 
}

\address{ S.N.Bose National Centre For Basic Sciences\\
Salt Lake City, Block JD, Sector III\\
Calcutta-700 091, INDIA\\
}

\author{Anita Mehta\footnote{
anita@boson.bose.res.in }}

\address{ 
S.N.Bose National Centre For Basic Sciences\\
Salt Lake City, Block JD, Sector III\\
Calcutta-700 091, INDIA\\
and\\
Institute of Theoretical Physics\\
University of California\\
Santa Barbara, CA 93106, USA
}
\author{J. K. Bhattacharjee\footnote{tpjkb@iacs.ernet.in}}

\address{ 
Department of Theoretical Physics\\
Indian Association for the Cultivation of Science\\
Jadavpur, Calcutta-700 032, INDIA}

\maketitle

\begin{abstract}
We present and analyse in this paper three models of coupled continuum
equations all united by a common theme: the intuitive notion
that sandpile surfaces are left smoother by the propagation
of avalanches across them. Two of these concern smoothing
at the `bare' interface, appropriate to intermittent avalanche
flow, while one of them models smoothing at the effective
surface defined by a cloud of flowing grains across the `bare'
interface, which is appropriate to the regime where avalanches
flow continuously across the sandpile.  
\end{abstract}

\draft
 


\section{INTRODUCTION}

The dynamics of sandpiles have
intrigued researchers in physics
over recent years \cite{kn:rpp,kn:ambook}
with a great deal of effort being devoted to
the development of techniques involving for instance
cellular automata \cite{kn:btw,kn:ca}, continuum
equations \cite{kn:mln,kn:amrjnsd,kn:bcre} and Monte Carlo schemes
\cite{kn:prl} to investigate this very complex
subject. However what have often been lost sight of
in all this complexity are some of the extremely
simple phenomena that are exhibited by granular media
which still remain unexplained.

One such phenomenon is that of the smoothing of
a sandpile surface after the propagation of an avalanche
\cite{kn:sdnagel}. It is clear what happens physically:
an avalanche provides a means of shaving off roughness
from the surface of a sandpile by transferring grains
from bumps to available voids \cite{kn:ambook,kn:ca}, and thus
leaves in its wake a smoother surface. However, surprisingly,
researchers have not to our knowledge come up with
models of sandpiles that have exhibited this behaviour.

In particular what has not attracted enough attention
in the literature is the qualitative difference
between the situations which obtain when
sandpiles exhibit intermittent and continuous
avalanches \cite{kn:degennes}. In this paper
we examine both the latter situations, 
via distinct models of sandpile surfaces.

A particular experimental paradigm that we
choose to put our discussions in context
is that of sand in rotating cylinders \cite{kn:frjohns,kn:nagel}. 
In the case when sand is
rotated slowly in a cylinder, 
intermittent avalanching is observed; thus
sand accumulates in part of the cylinder
to beyond its angle of repose \cite{kn:rmpnag}
and is then released via an avalanche
process across the slope. This happens
intermittently, since the rotation speed
is less than the characteristic time
between avalanches.
By contrast, when the rotation speed exceeds the time
between avalanches, we see continuous avalanching
on the sandpile surface.
Though this phenomenon has been observed
\cite{kn:rmpnag} and analysed physically
\cite{kn:degennes} in terms
of avalanche statistics, we are not aware of
measurements which measure the characteristics
of the resulting surface in terms
of its smoothness or otherwise.

What we focus on here 
is precisely this aspect, and make predictions
which we hope will be tested experimentally. 
In order to discuss this, we introduce first
the notion that granular dynamics is well
described by the competition between
the dynamics of grains moving independently
of each other and that of their collective
motion within clusters \cite{kn:ambook}.
A convenient way of representing this is via  coupled continuum equations
with a specific coupling between mobile grains $\rho$
and clusters $h$  on the surface of a sandpile \cite{kn:mln}. 
In the regime of intermittent avalanching,
we expect that the interface will be
the one defined by the `bare' surface,
$i.e.$ the one defined by the relatively immobile clusters
across which grains flow intermittently.
This then implies that the roughening 
characteristics of the $h$ profile 
should be examined. 
The simplest of the three models we discuss
in this paper (an exactly solvable model referred
to hereafter as Case A)
as well as the most complex one (referred to
hereafter as Case C) treat this situation,
where we obtain in both cases an asymptotic 
smoothing behaviour in $h$.
When on the other hand, there is continuous avalanching,
the flowing grains provide an effective film
across the bare surface and it is therefore
the species $\rho$ which should be analysed
for spatial and temporal roughening.
In the model hereafter referred to as Case B
we look at this situation, and obtain the surprising result
of a gradual crossover between purely
diffusive behaviour and  hypersmooth behaviour. 
In each case we present analytical results
pertaining to the continuum models
and compare the predictions so obtained
with the results obtained by numerical simulations
of the corresponding discretised equations. 

In general, the complexity of sandpile dynamics
leads us to equations which are coupled,
nonlinear and noisy: these equations present challenges
to the theoretical physicist in more ways than the obvious
ones to do with their detailed analysis and/or their
numerical solutions. 
 In particular,
our analysis of Case C reveals
 the presence of hidden length scales
whose existence was suspected analytically, but not demonstrated
numerically
in earlier work \cite{kn:mln,kn:pram}.

The normal procedure for probing temporal and spatial roughening
in interface problems is to  determine
the asymptotic behaviour of the interfacial
width with respect to time and space, via the
single Fourier transform. Here only
one of the variables, $(x,t)$ is integrated over in Fourier space,
and appropriate scaling relations are invoked to
determine the critical  exponents which govern this behaviour.
However, it turns out that this leads to ambiguities 
for those classes of problems where
there is an absence of simple scaling, or to be more
specific, where multiple length scales exist. In such
cases we demonstrate
that the double Fourier transform (where
{\it both} time and space are integrated over)
yields insights that are harder to obtain via the single
Fourier transform.

This point is illustrated by Case A, an exactly solvable model that
we introduce; we then use it to understand Case C, 
a nonlinear model where our analytical
results are clearly only approximations to the truth.

In order to make some of these ideas more concrete,
we now review some general facts about rough interfaces
\cite{kn:interf}.
Three critical exponents, $\alpha$, $\beta$, and $z$,
characterise the spatial and temporal scaling
behaviour of a rough interface.
They are conveniently defined by considering the (connected)
two-point correlation function of the heights
\be
S(x-x^{\prime},t-t^{\prime})=\bigl\langle 
h(x,t)h(x^{\prime},t^{\prime})\bigr\rangle
-\bigl\langle h(x,t)\bigr\rangle\bigl\langle h(x^{\prime},t^{\prime}
)\bigr\rangle.
\label{eq-strucdef}
\ee
We have
\[
S(x,0)\sim \vert x\vert^{2\alpha}\quad(\vert x\vert\to\infty)
\quad  \mbox{and} \quad S(0,t)\sim \vert t\vert^{2\beta}\quad(\vert t\vert
\to\infty),
\]
and more generally
\[
S(x,t)\approx\vert x\vert^{2\alpha}F
\bigl(\vert t\vert/\vert x\vert^z\bigr)
\]
in the whole long-distance scaling regime (${ x}$ and ${ t}$ large).
The scaling function $F$ is universal in the usual sense;
$\alpha$ and $z=\alpha/\beta$ are respectively referred to
as the roughness exponent and the dynamical exponent of the problem.
In addition, we have for the full structure factor which is the
double Fourier transform $S(k,\omega)$
\[
S(k,\omega) \sim \omega^{-1}k^{-1-2\alpha}\Phi(\omega/k^{z} )
\]
which gives
in the limit of small $k$ and $\omega$,
\be
S(k,\omega = 0) \sim k^{-1-2\alpha -z} \hskip 0.4cm ( k\rarrow 0)
  \quad \mbox{and} \quad
S(k=0,\omega) \sim \omega^{-1-2\beta - 1/z} \hskip 0.4cm ( \omega\rarrow 0)
\label{eq-kw1}
\ee
The scaling relations for the corresponding single Fourier transforms are
\be
S(k,t = 0) \sim k^{-1-2\alpha} \hskip 0.4cm ( k\rarrow 0) \quad \mbox{and} \quad
S(x=0,\omega) \sim \omega^{-1-2\beta} \hskip 0.4cm ( \omega\rarrow 0) \label{eq-1d}
\ee

In particular we note that the scaling relations
for $S(k,\omega)$ (Eq.(\ref{eq-kw1})) always 
involve the simultaneous presence
of $\alpha$ and $\beta$, whereas those corresponding
to $S(x,\omega)$ and $S(k,t)$ involve these exponents
{\it individually}. 
Thus, in order to evaluate the double Fourier transforms,
we need in each case information from the growing
as well as the saturated interface (the former being
necessary for $\beta $ and the latter for $\alpha$) whereas
for the single Fourier transforms, we need only 
information from the saturated interface for
$S(k,t = 0)$ and information from the growing interface for
$S(x=0,\omega)$. On the other hand, the information
that we will get out of the double Fourier transform
will provide a more unambiguous picture in the case
where multiple length scales are present, something
which cannot easily be obtained in every case with
the single Fourier transform. 

In Sections II, III and IV we present, analyse and discuss the
results of
Cases A, B and C respectively. 
Finally, in Section V, we reflect on the unifying features
of these models, and 
make some educated guesses on the dynamical
behaviour of real sandpile surfaces.

\section {CASE A: THE EDWARDS-WILKINSON EQUATION WITH FLOW}

Our first model involves
a pair of linear coupled equations, where the equation
governing the evolution of clusters (``stuck" grains) $h$
is closely related to the very well-known Edwards-Wilkinson (EW)
model \cite{kn:ew}. The equations are:
\begin{mathletters}
\begin{eqnarray}
{\pa h(x,t)\over \pa t}             & =& D_h\na^{2}h(x,t) +c\na h(x,t)+\eta(x,t)  
\label{eq-few}
\\
{\pa\rho(x,t)\over\pa t}            & =& D_{\rho}\na^{2}\rho(x,t) -c\na h(x,t)   
\label{eq-fewc}
\end{eqnarray}
\end{mathletters}
where the first of the equations describes the height $h(x,t)$
of the sandpile surface at $(x,t)$ measured from some mean $\langle h \rangle$,
and is precisely the EW equation in the presence of the flow
term $c\na h$. The second equation describes the evolution of
flowing grains, where $\rho(x,t)$ is the local density of such grains
at any point $(x,t)$. As usual, the noise $\eta(x,t)$ is taken
to be Gaussian so that:
$$ \lan \eta(x,t)\eta(x^{\pr},t^{\pr})\ran=\Delta^{2}\delta(x-x^{\pr})\delta(t-t^{\pr}).$$
with $\Delta $ the strength of the noise. Here, $\langle \cdots \rangle$ refers
to an average over space as well as over noise.

\subsection{Analysis of the decoupled equation in $h$}
For the purposes of analysis, we focus on the first of the
two coupled equations  (Eq.(\ref{eq-few})) presented above,
\[
{\pa h\over \pa t}= D_h\na^{2}h +c\na h+\eta(x,t)
\]
 noting that this equation 
is essentially decoupled from the second. (This statement
is, however, not true in reverse, which has implications
to be discussed later). 
We note that this is entirely equivalent to the
 Edwards-Wilkinson equation \cite{kn:ew} in a frame moving with velocity $c$
$$ x^{\pr} = x+ct \, , \quad  t^{\pr} = t $$
and would on these grounds expect to find only the
well-known EW exponents $\alpha=0.5$ and $\beta=0.25$ \cite{kn:ew}.
This would be verified by naive single Fourier transform analysis 
of Eq.(\ref{eq-few}) which yields these exponents via 
Eq.(\ref{eq-1d}).

 Equation (\ref{eq-few}) 
can be solved exactly as follows. 
The propagator $G(k,\omega)$ is
$$ G_h(k,\omega) = (-i\omega+D_h k^2+i kc )^{-1} $$
This can be used to evaluate the structure factor
\[  
S_h(k,\omega) = {\langle h(k,\omega)h(k^{\pr},\omega^{\pr}) \rangle \over 
\delta(k+k^{\pr})\delta(\omega+\omega^{\pr}) }
\]
which is the Fourier transform of the full correlation function
$S_h(x-x^{\pr},t-t^{\pr})$ defined by Eq.(\ref{eq-strucdef}).
The solution for $S_h(k,\omega)$ so obtained is: 
\be
S_h(k,\omega)  =  {\Delta^2 \over (\omega-ck)^{2}+D_h^{2}k^{4}} 
\label{eq-skw_A}
\ee

This is illustrated in Fig.\ref{fig-A1} while 
representative graphs for 
 $ S_h(k,\omega=0)$ and 
 $S_h(k=0,\omega)$ are presented in Figs.\ref{fig-A2} and \ref{fig-A3} respectively.
Before proceeding further, we make  the following observation about
the double Fourier transform $S_h(k=0,\omega)$; this shows
an $\omega^{-2}$
behaviour coming from 
Eq.(\ref{eq-skw_A}), which we will also see later.
We mention here that the ubiquity of this
$\omega^{-2}$
arises from the form of the scaling relation
Eq.(\ref{eq-kw1}),
which is relevant
for frequencies $\omega < \omega_c \approx k^{z_h}$, whereas for
$\omega>\omega_c$ the high frequency behaviour 
takes over giving $\omega^{-2}$ (cf. Eq.(\ref{eq-delta}) in the Appendix).
As $k=0$ for the purposes of calculation of this structure factor,
it is always the high frequency behaviour that dominates,
leading to the ubiquity of $\omega^{-2}$ whenever it is measured. \\

It is obvious from Eq.({\ref{eq-skw_A})  that $S_h(k,\omega)$ does not show simple scaling. 
More explicitly, if we write 
$$ S^{-1}_h(k,\omega=0) = \frac{\omega_0^2}{\Delta^2} \left( \frac{k}{k_0} \right)^2
\left[ 1 + \left( \frac{k}{k_0} \right)^2 \right]$$ 
with $k_{0} = c/ D_h$, 
 and $\omega_0 = c^2/D_h$, we see that there are 
two limiting cases :

\begin{itemize}
\item for $k \gg  k_{0}$,  
 $ S^{-1}_h(k,\omega=0) \sim k^{4}$; using again 
 $S^{-1}_h(k=0,\omega) \sim \omega^{2}$,
we obtain   $\alpha_h=1/2$ and $\beta_h=1/4$, $z_h=2$ via
Eqs.(\ref{eq-kw1}).  
\item for $k \ll k_{0}$, $ S^{-1}_h(k,\omega=0) \sim
k^{2}$; using the fact that
the limit $S^{-1}_h(k=0,\omega)$ is always $\omega^{2}$, this
is consistent with the set of exponents
 $\alpha_h=0 $, $\beta_h=0$ and $z_h=1$ via
Eqs.(\ref{eq-kw1}). 
\end{itemize}

The first of these contains no surprises, 
being the normal EW fixed point
\cite{kn:ew},
while the second represents a new,
 `smoothing' fixed point.  

We now explain this smoothing fixed point via a 
simple physical picture. 
The competition between the two terms in Eq.(\ref{eq-few})
determines the nature of the fixed point observed:
when the diffusive term dominates the flow term,
the canonical EW fixed point is obtained, in the limit
of large wavevectors $k$. On the contrary, when the
flow term predominates, the effect of diffusion
is suppressed by that of a travelling wave
whose net result is to penalise large slopes; this
leads to the smoothing fixed point obtained
in the case of small wavevectors $k$.
We emphasise however, that this is a toy model
of smoothing, which will be used to illuminate
the  discussion of models B and C below. 

\subsection{Coupled equations: a model of smoothing}
 
We realise from the above that the interface $h$
is smoothed because of the action of the  flow term which 
penalises the sustenance of finite gradients $\na h$ in 
Eq.(\ref{eq-few}). However, Eq.(\ref{eq-few}) is effectively decoupled 
from Eq.(\ref{eq-fewc}), while Eq.(\ref{eq-fewc}) is manifestly 
coupled to Eq.(\ref{eq-few}). 
In order for the coupled Eqs.(4) to qualify 
as a valid model of sandpile dynamics, we would need to ensure that 
no instabilities are generated in either of these  
by the coupling term  $c\na h$.

In this spirit, we look first at the value of $\rho$ 
averaged over the sandpile, as a function of time 
(Fig.\ref{fig-A6}a). We observe that the incursions 
of $\langle\rho\rangle$ 
into negative values are limited to relatively 
small values, suggesting
that the addition of a constant background 
of $\rho$ exceeding this negative value would 
render the coupled system meaningful, at least to a first approximation.
In order to ensure that this average does 
not involve wild fluctuations, 
we examine the fluctuations in $\rho$,
viz.$ \sqrt{\langle\rho^{2}\rangle - \langle\rho\rangle^{2}}$ 
(Fig.\ref{fig-A6}b). The trends in that
figure indicate that this quantity appears to saturate,
at least upto computationally accessible times.
Finally we look at the {\it minimum} and {\it maximum} value of
$\rho $ at any point in the pile over a large range
of times (Fig.\ref{fig-A6}c);  this appears to be
bounded by a modest (negative) value of `bare' $\rho$.
Our conclusions are thus that the 
fluctuations in $\rho$ saturate at computationally accessible
 times  and that the
negativity of the fluctuations in $\rho $ can always
be handled by starting with a constant $\rho_0$, a
constant `background' of flowing grains, which is
more positive than the largest negative fluctuation.

Physically, then, the above implies that at 
least in the presence of a constant large density $\rho_0$ of
flowing grains, it is possible to induce the level 
of smoothing corresponding
to the fixed point $\alpha = \beta = 0.$ This model is
thus one of the simplest possible ways in which
one can obtain a representation of 
the smoothing of the `bare surface' that is frequently observed
in experiments on real sandpiles after intermittent avalanche 
propagation \cite{kn:sdnagel}.

\section {CASE B: A SIMPLE FORM OF COUPLING, WITH COMPLEX CONSEQUENCES }

Our model equations, first presented
in \cite{kn:mln} involve a simple coupling
between the species $h$ and $\rho$,
where the transfer between the species occurs
only in the presence of the flowing grains
and is therefore relevant to the regime
of continuous avalanching when the duration
of the avalanches is {\it large} compared to the time
between them. The equations are:
\begin{mathletters}
\begin{eqnarray}
\label{eq-cup}
{\pa h(x,t) \over \pa t}    &=& D_h\na^{2} h(x,t) -T(h,\rho) + \eta_{h}(x,t) \\
{\pa \rho(x,t) \over \pa t} &=& D_{\rho}\na^{2} \rho(x,t) +T(h,\rho) +\eta_{\rho}(x,t) \\
T(h,\rho) &=& -\mu\rho({\na h)} 
\end{eqnarray}
\end{mathletters}
where the terms $\eta_h(x,t) $ and $\eta_{\rho}(x,t) $
represent  
Gaussian white noise as usual:
\begin{eqnarray}
\langle \eta_h(x,t)\eta_h(x^{\pr},t^{\pr}) \rangle  
& = & \Delta_h^2\delta(x-x^{\pr})\delta(t-t^{\pr}) \nonumber\\
\langle \eta_\rho(x,t)\eta_\rho(x^{\pr},t^{\pr}) \rangle 
& = & \Delta_\rho^2\delta(x-x^{\pr})\delta(t-t^{\pr}) \nonumber
\end{eqnarray}
and the $\langle \cdots \rangle $ stands for average over space as well 
as noise.

A simple physical
picture of the coupling or `transfer' term $T(h,\rho)$ between
$h$ and $\rho$ is the following: flowing grains
are added in proportion to their local density to 
regions of the interface
which are at less  than the critical slope, and vice versa,
{\it provided that the local density of flowing grains
is always non-zero}. 
This form of interaction becomes zero in
the absence of a finite density of flowing
grains $\rho$ (when the equations become
decoupled) and is thus the simplest form  appropriate
to the situation of continuous avalanching
in sandpiles.
We analyse in the following the
profiles of $h$ and $\rho$ consequent on this form.

It turns out that a singularity 
discovered by Edwards \cite{kn:edw}
three decades ago in the context of fluid turbulence is present in 
models with a particular form of the transfer term $T$ ; the above is
one example, while   
another example is  the model due to Bouchaud {\it et al.} (BCRE)
\cite{kn:bcre} where
\[ T = -\nu \na h  - \mu \rho (\na h) \]
and the noise is present only in the equation of motion for $h$. 
This singularity, the so-called infrared divergence,
largely controls the dynamics and produces unexpected exponents.

\subsection{Theoretical analysis}

We carry out first the theoretical analysis
of Eqs.(6). An examination of the above equations reveals the
presence of two likely length scales in each, one associated
with the diffusive motion, and the other with the so-called
transfer term $T(h,\rho)$, representing the coupling between the two
species. In these circumstances, a renormalisation group 
analysis would clearly be inappropriate due to the breakdown
of simple scaling. In recent years, however, a self-consistent mode
coupling analysis used hitherto in dynamic critical phenomena  \cite{kn:kawasaki}
has been used to look at in particular the Kardar-Parisi-Zhang (KPZ) 
equation \cite{kn:kpz,kn:doherty}
and we extend its use to the case of the coupled equations presented here. 

In this method we set up equations (to one-loop order) for the correlation 
functions and self-energies in terms of the full Green's functions,
correlation functions
and vertices using assumed scaling forms for each. 
The critical exponents $\alpha$ and $\beta$ defined above are obtained from
the self-consistent solutions of these equations using $D_h = D_\rho$.

Focusing on the $h$ variable to start with,
we define the Green's functions and the correlation functions of the $h$ and $\rho$
variables 
\begin{eqnarray} 
G_h(k,\omega)   &= & \left\langle {\delta h(k,\omega)\over 
\delta \eta(k^{\pr},\Omega)}
\right\rangle {1\over \delta(k+k^{\pr}) \delta(\omega+\Omega)} \nonumber\\
G_\rho(k,\omega)   &= & \left\langle {\delta \rho(k,\omega)\over 
\delta \eta(k^{\pr},\Omega)}
\right\rangle {1\over \delta(k+k^{\pr}) \delta(\omega+\Omega)} \nonumber\\
S_h(k,\omega)   &= & {\langle h(k,\omega)h(k^{\pr},\Omega)\rangle 
\over \delta(k+k^{\pr})\delta(\omega+\Omega)} \nonumber\\
S_\rho(k,\omega)   &= & {\langle \rho(k,\omega)\rho(k^{\pr},\Omega)\rangle 
\over \delta(k+k^{\pr})\delta(\omega+\Omega)} \nonumber
\end{eqnarray}

  The analysis of these functions will be in terms of a weak scaling hypothesis
which states 
\begin{eqnarray}
  G_h(k,\omega) = k^{-z_h} f_h(\frac{\omega}{k^{z_h}},\frac{\omega}{k^{z_\rho}})\nonumber\\
  G_\rho(k,\omega) = k^{-z_\rho} f_\rho(\frac{\omega}{k^{z_h}},\frac{\omega}{k^{z_\rho}})\nonumber
\end{eqnarray}
 A strong scaling would imply the existence of a single time scale $i.e.$ $z_h
= z_\rho$. As we show below, this cannot be the case here. The absence of strong 
scaling implies that the roughness exponents $\alpha_h$ and $\alpha_\rho$ may 
become functions of $k$.

We consider the full Green's function $G_h(k,\omega)$, which is
given via the well-known Dyson equation \cite{kn:mahan} ,
\[
G_h^{-1}(k,\omega) = {G_h^{0}}^{-1}(k,\omega) + \Sigma_h(k,\omega)
\] 
Here, the zeroth order Green's function is 
\[
G_h^{0}(k,\omega) = (-i\omega+k^2)^{-1} \nonumber
\]
The scaling forms of the functions $G_h(k,\omega)$ and $S_h(k,\omega)$ are 
given by, in the limit $\omega \rarrow 0$,
\begin{eqnarray}
 G_h(k,\omega)
     & \sim & \frac{1}{i\omega + k^2 + k^{z_h}} \nonumber \\
 S_h(k,\omega) & \sim & \frac{1}{k^{1 + 2\alpha_h - z_h}} 
   \left( \frac{1}{\omega^2 + k^{2z_h}} \right) \nonumber
\end{eqnarray}
Similar scaling relations hold for the species $\rho$.

To one-loop order, the self-energy $ \Sigma_h(k)$ is given by (Fig.\ref{fig-fd1}b)
\begin{mathletters}
\begin{eqnarray}
\Sigma_h(k,\omega) & = & \mu^{2} \int {dq\over 2\pi}\int {d\Omega\over 2\pi}
G_h(k-q,\omega-\Omega)S_{\rho}(q,\Omega) \, k(k-q) \label{eq-seh1} \\
& \sim & \mu^2  \int {dq\over 2\pi}\int {d\Omega\over 2\pi} 
\left[ \frac{1}{i(\omega - \Omega) + \Sigma_h(k - q,\omega-\Omega)}\right] 
\frac{k(k-q)}{q^{1 + 2\alpha_\rho}} 
\left[ \frac{2\Sigma_\rho(q,\Omega)}{\Omega^2 + |\Sigma_\rho(q,\Omega)|^2}  
\right]\label{eq-seh2}
\end{eqnarray}
\end{mathletters}
where the second line follows from the first in the limit of small $\Omega$. 
We note that due to the presence of the term $q^{-1-2\alpha_\rho}$, the
integral is dominated by the 
singularity in the integrand at $q \rightarrow 0 $. This `infrared divergence'
which results from  the divergence of the {\it internal} momenta $q$, 
is very different from the usual divergences encountered
in critical phenomena where the latter occur 
for small wave numbers and are associated with
long wavelength instabilities in the external momenta. In this case due to  
the infrared divergence in the above equation in the internal momenta $q$,
the integral diverges {\it for any value of the external momenta $k$},
 so long as $\alpha_{\rho}>0$.

We thus need either to evaluate the integral with a lower cut-off $k_0$ or to
introduce a suitable regulator. We follow the first of these procedures
for the above equation, and the second of the 
procedures to do with the corresponding quantity, $S_{\rho}(k,\omega)$, for $\rho$.

We then proceed to evaluate the self-energy at zero external frequency, $i.e$
$\Sigma_h ( k, \omega=0) $ from Eq.(\ref{eq-seh1}).
As $q \rarrow 0 $ we can approximate $G_h(k-q,-\Omega)$ by 
\begin{eqnarray}
G_h^{-1}(k,-\Omega) & =  & i\Omega + k^2 + \Sigma_h(k,-\Omega)\nonumber \\
 & \approx & k^2 + \Sigma_h(k,0) \nonumber
\end{eqnarray}
where the second line follows from the fact that we are looking at the 
 $q \simeq 0$ limit of the internal frequency $\Omega \sim q^{z_h} $. 
As $\Sigma_h(k,0) \sim k^{z_h}$, the small $k$ behaviour of $G_h(k)$ is 
dominated by $\Sigma_h(k)$ for $z_h < 2$ , $i.e$
\[ G_h^{-1}(k) \sim \Sigma_h(k) \]
The integral in Eq.(\ref{eq-seh1}) becomes in the limit of zero external 
frequencies  
$$ \Sigma_h(k) = {\mu^{2}k^{2}\over \Sigma_h(k)} \int {dq \over 2\pi}
\int {d\Omega \over 2\pi} S_{\rho}( q,\Omega)  $$
Using the scaling form for the single Fourier transform (Eq.(\ref{eq-1d}))
we find
\[
\Sigma_h(k) = \mu^{2}k^{2} {\Sigma_h(k)}^{-1}C_{\rho} \int {dq \over 2\pi}{1\over q^{1+2\alpha_{\rho}}}  
\]
We now have to evaluate the integral by cutting off the momentum integration
at $k_{0} \ll 1$ , $i.e.$ we follow the first of the procedures
given above to handle the infrared divergence. This gives, after
some simplification,
\[
{\Sigma_h^{2}(k)}= \mu^{2}k^{2} {k_{0}^{-2\alpha_{\rho}}
C_{\rho}\over 4\pi\alpha_\rho} 
\] 
From the above equation with the scaling 
relation $\Sigma_h(k) \sim k^{z_h} $
we find, on equating powers of $k$,
$$ z_h = 1$$

We note here that the presence 
of the term $\rho\na h$ 
could in principle cause the
vertex $\mu$ to renormalise,
leading to a correction to 
$ z_h$.
In these circumstances, the
expression for the self-energy $\Sigma_h(k,\omega=0)$  is given by 
\be
\Sigma_h(k,\omega=0) =\mu^{2} \int {dq\over 2\pi}\int {d\Omega\over 2\pi}\Gamma_3(k,q,k-q)G_h(k-q,-\Omega)
S_{\rho}(q,\Omega) k(k-q)
\ee 
where we have introduced a three-point vertex function $\Gamma_3(k,q,k-q)$ in
Eq.(\ref{eq-seh1}). 
Assuming that as $ q\rarrow 0$, we can write the asymptotic form
for the three-point vertex as,
\be
\label{eq-xmu}
\Gamma_3(k,q,k-q) \sim k^{x_{\mu}} 
\ee
we find 
$$  z_h = 1+ {x_{\mu}\over 2} $$
In the event that numerical results suggest
$ z_h \ne 1$
we will have to incorporate this new renormalised
vertex into our calculations.

Next we examine the correlation function for $h$, $S_h(k,\omega)$, which
to one-loop order is given by (Fig.\ref{fig-fd2}a)
\begin{mathletters}
\begin{eqnarray}
S_h(k,\omega) & = & \frac{1}{\omega^2 + |\Sigma_h(k,\omega)|^2}
 \left[ 1 + \mu^2\int{dq\over 2\pi}\int{d\Omega\over 2\pi}  
 |k-q|^2 S_h(k-q,\omega-\Omega) S_\rho(q,\Omega)\right]\\
& \approx & \frac{1}{\omega^2 + |\Sigma_h(k,\omega)|^2}
\left[ 1  + \mu^2\int{dq\over 2\pi}\int{d\Omega\over 2\pi} 
 \frac{|k-q|^2}{|k-q|^{1+2\alpha_h}}
 \frac{1}{q^{1+2\alpha_\rho}}
 \left( \frac{2\Sigma_\rho(q,\Omega)}{\Omega^2 + |\Sigma_\rho(q,\Omega)|^2}  \right)
 \right. \nonumber \\
 & & \left. \hspace{70mm}
\left( \frac{2\Sigma_h(k-q,\omega-\Omega)}{(\omega-\Omega)^2 + |\Sigma_h(k-q,\omega-\Omega)|^2} \right)
\right]\\
& \approx & \frac{1}{\omega^2 + |\Sigma_h(k,\omega)|^2}\left[ 1  +  \mu^2\int{dq\over 2\pi}
 \frac{|k-q|^{1-2\alpha_h}}{q^{1+2\alpha_\rho}} 
 \left(
 \frac{\Sigma_\rho(q) + \Sigma_h(k-q)}{\omega^2 + \left(\Sigma_\rho(q) + \Sigma_h(k-q)\right)^2}\right)
\right]\label{eq-shkw}
\end{eqnarray}
\end{mathletters}
The frequency-dependent self-energy
 $\Sigma_h(k,\omega)$
 in the above is given by 
evaluating the integral over the internal frequency $\Omega$ in
Eq.(\ref{eq-seh2}). This leads to  
\begin{mathletters}
\begin{eqnarray}
 \Sigma_h(k,\omega)  & \approx & \mu^2\int\frac{dq}{2\pi} \frac{k(k-q)}{q^{1+2\alpha_\rho}}
      \frac{A}{-i\omega + \Sigma_\rho(q) + \Sigma_h(k-q)}\\
  & \approx & \mu^2\frac{A}{4\pi\alpha_\rho} \frac{k^2 k_0^{-2\alpha_\rho}}{-i\omega + \Gamma_0 k}\\
  & \approx & \frac{\Gamma_0^2 k^2}{-i\omega + \Gamma_0 k} 
\end{eqnarray}
\end{mathletters}
where $\Gamma_0 =\mu k_0^{-\alpha_\rho}\sqrt{\frac{A}{4\pi\alpha_\rho}}$,
and the second line in the above follows from taking a 
$q\rightarrow 0$
limit and introducing a cutoff wavevector $k_0$
in the integral on the first line.
Introducing this expression for 
 $\Sigma_h(k,\omega)$
in Eq.(\ref{eq-shkw}) and recognising that  
the divergence due to  $q^{-(1+2\alpha_\rho)}$ dominates the integral
we find
\begin{eqnarray}
S_h(k,\omega) & = & \left(\omega^2+ \frac{\Gamma_0^4 k^4}{\omega^2 + \Gamma_0^2 k^2}\right)^{-1}
 \left[ 1 + \frac{\mu^2C_\rho}{4\pi\alpha_\rho}
k_0^{-2\alpha_\rho}\frac{k^2}{k^{1+2\alpha_h}}
\frac{\Gamma_0k}{\omega^2+\Gamma_0^2k^2} \right]\label{eq-shkw2}
\end{eqnarray}
On integrating with respect to $\omega$ we can write the structure factor $S_h(k,t=0)$ as
\be S_h(k,t=0) \equiv \int S_h(k,\omega) {d\omega \over 2\pi} = \frac{A_0}{k} + \frac{B_0}{k^{1+
2\alpha_h}} \label{eq-shkt} \ee

Recognising that  the scaling form of $S_h(k,t=0) \sim k^{-1-2\alpha_h}$, we
notice that $\alpha_h$ cannot in general be determined  from  Eq.(\ref{eq-shkt}).
This is because the second term on the right-hand-side of Eq.(\ref{eq-shkt})
dominates at small momenta $k$ provided $\alpha_h > 0 $, 
indicating that $\alpha_h$ is indeterminate to this order of calculation.  

We turn now to the critical exponents
in $\rho$. The single loop 
self-energy $\Sigma_{\rho}(k,\omega)$ is given as shown in Fig.\ref{fig-fd1}a
by
\be
\Sigma_{\rho}(k,\omega=0) = -\mu^{2}\int {dq \over 2\pi}
\int {d\Omega \over 2\pi} G_{\rho}(k-q,-\Omega)S_h(q,\Omega) q^{2} 
\ee
Inserting the expressions for $G_\rho(k-q,\omega-\Omega)$ and $S_h(q,\Omega)$
 we find
\[
\Sigma_{\rho}(k,\omega=0)  =  -\mu^{2}\int {dq \over 2\pi} \int {d\Omega \over 2\pi}
\left[{1\over {i \Omega+ |k-q|^{z_\rho}}}\right]
\left[{2q^{z_h}\over \Omega^2 +  q^{2z_h}} \right]{q^2\over q^{1+2\alpha_h}} 
\]
This gives, on performing the integral over internal frequency $\Omega$,
\[
\Sigma_{\rho}(k,\omega=0) = -\mu^{2}\int {dq \over 2\pi}{q^2 \over q^{1+2\alpha_h}}
{1\over {\vert k-q \vert}^{z_\rho} + q^{z_h}}
\]
In order to discuss
this further in the context of $z_\rho$, we need to make
a statement about  $\alpha_h $ 
and $z_h$. We have already obtained  
 $z_h = 1$ in the foregoing 
and will now quote our numerical result for
 $\alpha_h$, viz. 
 $\alpha_h=0.5 $.
For small $k$ the self-energy can then be written as 
\[
  \Sigma_\rho(k,\omega) \simeq -\mu^2  \left[
   \int \frac{dq}{2\pi} \frac{1}{\left(q + q^{z_\rho}\right)}
  + z_\rho k  \int \frac{dq}{2\pi} \frac{1}{\left(q + q^{z_\rho}\right)
    \left( q + q^{2z_\rho}\right)} \right]
\]
We see from the above that
$\Sigma_\rho(k,0)$, the relaxation rate for $\rho$ fluctuations, is 
negative and finite as $k \rightarrow 0$, 
 and we  need to add a positive constant, $\Sigma_0$, to the
self-energy ($\Sigma_0 > |\Sigma_\rho(k \rightarrow 0)|$) for regulatory purposes. 
This divergence in the relaxation rate, needing
regulation, is reflected in the divergence we have encountered
in our numerical investigations below;  we have there
followed an analogous procedure by introducing a numerical
regulator which replaces divergent values of the transfer
term by suitably defined cutoffs \cite{kn:mln}.
The resulting constancy of $\Sigma_{\rho}$ implies $z_{\rho} \approx 0 $ 
for the regulated equations and will be used in the following.
  
The correlation function $S_{\rho}(k,\omega)$ is given by (Fig.\ref{fig-fd2}b)
\begin{eqnarray}
S_{\rho}(k,\omega) & = & {1\over (\omega^{2}+k^{2z_\rho})} \int {dq\over 2\pi} 
\int{d\Omega\over 2\pi}(k-q)^2 S_h(k-q,\omega-\Omega)S_\rho(q,\Omega) \nonumber 
\end{eqnarray} 

The above integral will now be evaluated in the limit 
$ q \rarrow 0 $ and since $\Omega \sim q^{z_h} $ for $S_h$ 
we can replace $$S_h(k-q,\omega-\Omega) \simeq S_h(k,\omega). $$ 
Then using the scaling relation Eq.(\ref{eq-1d}) we have 
\begin{mathletters}
\begin{eqnarray}
 S_\rho(k,\omega) & \simeq & {1\over (\omega^{2}+ k^{2z_\rho})}\int {dq\over 2\pi}{C_\rho  \over q^{1+2\alpha_\rho}} k^{2} S_h(k,\omega)  \\
& = & {k^{2}C_\rho \over (\omega^{2}+k^{2z_\rho})} S_h(k,\omega)\int {dq \over 2\pi}{1\over q^{1+2\alpha_\rho}}  \\
& = & {C_\rho k_0^{-2\alpha_\rho} \over 4\pi \alpha_\rho} 
{k^{1-2\alpha_h+z_h} \over (\omega^{2} +k^{2z_{\rho}})(\omega^2+k^{2z_h})}\label{eq-srkw}
\end{eqnarray}
\end{mathletters}
where the last step follows from introducing a lower cutoff $k_0$ 
in the momentum integration over $q$. 

Using Eq.(\ref{eq-1d})
we have after integrating Eq.(\ref{eq-srkw}) over $\omega$ 
\begin{equation}
 S_\rho(k,t=0) \sim k^{-(1+2\alpha_\rho)}  \sim  
{k^{1-2\alpha_h} \over k^{z_\rho} ( k^{z_h}+k^{z_\rho} )}
\end{equation}
Finally using $z_{\rho} \approx 0 $ we have 
\begin{mathletters}
\begin{eqnarray}
\alpha_{\rho} = \alpha_h+ {z_h\over 2}-1 & \quad & \mbox{for large $k$}\label{eq-rho1} \\
\alpha_{\rho} = \alpha_h-1  & \quad & \mbox{for small $k$}\label{eq-rho2} 
\end{eqnarray}
\end{mathletters}
Given our numerical result of $\alpha_h = 0.5$,
the above predicts a negative $\alpha_\rho$,
at small $k$.
This is consistent with, and validates
our assumption of, a cutoff $k_0$
which arises naturally as the wavevector
separating the region of $\alpha_\rho < 0$
(no infrared divergence) and
$\alpha_\rho > 0$ (infrared divergence prevalent)
in Eqs.(\ref{eq-seh2}) and (\ref{eq-shkw}).

More importantly, this non-trivial result for $\alpha_{\rho}$
indicates that should we see numerical evidence of a negative
$\alpha_{\rho}$ for small wavevectors, we will have verified
the existence of an asymptotic hypersmoothing in our
model equations, which has an important bearing
on sandpile surfaces in the continuous avalanching regime.
This is discussed further in our concluding section.
  
\subsection{Numerical Analysis}

We focus now on our numerical
results for Case B.
The coupled equations in this section and the following 
one were numerically integrated by using the method of finite differences
\cite{kn:gears}.
Our grids in time and space were kept as fine-grained as
computational constraints allowed so that our grid size
in space $\Delta x$ was chosen to be in the range (0.1,0.5) 
whereas that in time was in the range
$\Delta t $ (0.001, 0.005). Thus the
instabilities associated
with the discretisation of nonlinear continuum equations 
were avoided and convergence was checked 
by keeping $\Delta t $ small enough such that the quantities under
investigation were independent
of further discretisation. Our results were also checked for 
finite size effects.
In the calculations of this section we chose $ D_h =
D_{\rho} = 1.0$
and $\mu$ = 1
and our results were 
averaged over several independent configurations.
We have calculated the exponents 
$\alpha$ and $\beta$ and the corresponding error 
bars using the linear least square 
fit so that 
$-(1+2\beta)$ and $-(1+2\alpha)$ are given by slopes 
of the fitted straight lines.

On discretising the equations Eqs.(6)
we found once again the divergences
that were previously observed in \cite{kn:mln}.
These divergences are in our
view a direct representation
of the infrared divergence
mentioned above, and we follow
here a parallel course in regulating
these via an explicit regulator.
In earlier work \cite{kn:mln}, a regulator
was introduced which replaced
the function $\mu\rho\na h$ 
by the following:
\begin{eqnarray}
  T &=&  +1  \quad\quad\quad\mbox{for}\quad \mu \rho (\na h) > 1 \nonumber \\
    &=&  \mu \rho (\na h)   \quad \mbox{for} \quad -1\le \mu\rho(\na h) \le 1 \nonumber\\
    &=&  -1  \quad \quad\quad\mbox{for} \quad \mu \rho (\na h) < -1  \nonumber
\end{eqnarray}
In addition in this paper,
we have introduced noise reduction
to the regulated equations
which has led to a more accurate
evaluation of all our critical
exponents.

The Fourier transform 
$ S_h(k,t=0) $ (Fig.\ref{fig-B1}) 
is consistent with
a spatial roughening exponent $\alpha_h \sim 0.501\pm 0.007   $  
via  our observation of
$$ S_h(k,t=0) \sim  k^{-2.03\pm 0.014 } $$
and the Fourier transform
$ S_h(x=0,\omega) $ (Fig.\ref{fig-B2}) 
is consistent with
a temporal roughening exponent 
$\beta_h \sim 0.465\pm 0.008 $
via  our observation of
$$ S_h(x=0,\omega) \sim {\omega^{-1.93\pm 0.017}} $$
Hence
$ z_{h}\sim 1.07$, and thus the exponent
$x_\mu \simeq 0$ (Eq.(\ref{eq-xmu})), indicating that
the $\mu$ vertex does not renormalise.

Using  $\alpha_h \sim 0.5 $ in Eq.(\ref{eq-shkw2}) we
can write the    
structure factor
$S_h(k,\omega)$ 
 as
\be
S_h(k,\omega) 
        =   \frac{1}{1+\Omega^2(1+\Omega^2)} 
   \left[ \frac{1 + \Omega^2}{\Gamma_0^2 k^2} + \frac{1}{\Gamma_0 k^3} \right]
 \label{eq-fit} 
\ee
where $\Omega= {\omega}/{\Gamma_0 k}$. 
We find from the above that the expected form of 
 $S_h(k,\omega=0) $ in the limit of small wavevectors to be 
\be 
S_h(k,\omega=0) \sim k^{-3} \label{eq-k3}
\ee
Realising that our computed $\alpha_h < 1$,
we obtain from Eq.(\ref{eq-shkw2}) the prediction
\be 
S_h(k=0,\omega) \sim \omega^{-2} \label{eq-hw2}
\ee

The full structure factor $S_h(k,\omega)$ has been 
calculated at two different $k $ points and  Fig.\ref{fig-B3}   
displays our results to Eq.(\ref{eq-fit}). 
The solid and the dashed line in the Fig.\ref{fig-B3} are the plots
of the Eq.(\ref{eq-fit}) for $k=0.1$ and $k=0.2$ 
with $\Gamma_0$ = 0.4 and 0.5 respectively.
The spatial structure factor
$ S_h(k,\omega=0)$ shows a power-law
behaviour (Fig.\ref{fig-B4}) given by 

 $$ S_h(k,\omega=0) \sim {k^{-3.40 \pm .029}} $$

in qualitative accord with Eq.(\ref{eq-k3}),
and the temporal structure factor
 $ S_h(k=0,\omega)$ shows a power-law
behaviour (Fig.\ref{fig-B5}) given by 

$$ S_h(k=0,\omega) \sim {\omega^{-1.91\pm .017}} $$

in accord with Eq.(\ref{eq-hw2}).

Given our values of $\alpha_h \simeq 0.5$ and $z_h \simeq 1$,
Eqs.(\ref{eq-rho1}) and (\ref{eq-rho2}) 
predict a crossover in $\alpha_\rho$ from 0 at large $k$ to
-0.5 as $k \rarrow 0$.
The single Fourier transform $S_\rho(k,t=0)$
(Fig.\ref{fig-B7})  
shows a crossover behaviour from 
$$ S_\rho(k,t=0)\sim {k^{-2.12\pm 0.017}} $$ 
for large wavevectors to
$$ S_\rho(k,t=0)\sim \mbox{constant} $$ 
as $k \rarrow 0$. 
 In  Fig.\ref{fig-B7} we find a crossover from 0.56 at large $k$
to -0.5 as $k \rarrow 0$, which shows the same trend as the prediction above.
Note however that the simulations
also manifest in addition to the above 
the normal diffusive behaviour represented
by $\alpha_\rho = 0.56 $ at large wavevectors.
The single Fourier transform in time $S_\rho(x=0,\omega)$ (Fig.\ref{fig-B6}) 
shows a power-law behaviour 
$$ S_\rho(x=0,\omega) \sim {\omega^{-1.81 \pm 0.017}} $$ 

While the range of wavevectors in Fig.15  over which crossover in
$ S_\rho(k,t=0)$ is observed
was restricted by our computational
constraints,
 the form of the crossover appears
 conclusive. 
Checks (with fewer averages) over larger system sizes revealed the same trend;
additionally our theoretical calculations support the observed
crossover via Eqs. (17).

\subsection {Homing in on the physics: a discussion of smoothing}
We focus in this section
on the physics of the equations
and our results.
In the regime of continuous avalanching
in sandpiles, the major dynamical
mechanism
is that of mobile grains $\rho$
present in avalanches
flowing into voids in the $h$ landscape
as well as the converse process
of unstable clusters (a surfeit of $\nabla h$
above some critical value) becoming
destabilised and adding to the avalanches.
Our results for the critical
exponents in $h$ indicate no further spatial
smoothing beyond the diffusive; however,
those in the species $\rho$ indicate
a crossover from purely diffusive
to an asymptotic hypersmooth behaviour.
Our claim for continuous avalanching is as follows:
{\it the flowing grains play
the major dynamical role
as all exchange between $h$ and
$\rho$ takes place only in the
presence of $\rho$. These
flowing grains therefore
distribute themselves over
the surface filling in voids
in proportion both to their local
density as well as to the depth
of the local voids; it is this
distribution process that leads
in the end to a strongly smoothed
profile in $\rho$.}
Additionally, since in the regime
of continuous avalanches, the effective interface
is defined by the profile of the {\it flowing} grains,
it is this profile that will be measured
experimentally for, say, a rotating
cylinder with high velocity of rotation.

\section{Anomalous smoothing: the case of tilt and boundary-layer exchange \\(Case C)}

The last case we discuss in this paper 
involves a more complex coupling between the
the stuck grains $h$ and the flowing grains $\rho$
as follows
\begin{mathletters}
\begin{eqnarray}
{\pa h(x,t) \over \pa t}    & = & D_h\na^{2} h(x,t) -T + \eta(x,t) \\
{\pa \rho(x,t) \over \pa t} & = & D_{\rho}\na^{2} \rho(x,t) +T \\
T(h,\rho) &=  & -\nu(\na h)_{-}-\lambda \rho(\na h)_{+} 
\label{eq-caseC}
\end{eqnarray}
\end{mathletters}
with $\eta(x,t) $ representing white noise as usual. 

Here, 

\begin{mathletters}
\begin{eqnarray}
z_+ &=&  z \quad \mbox{for}\quad z> 0 \nonumber\\
    &=&  0 \quad \mbox{otherwise}    \label{eq-step1}\\
z_- &=&  z \quad \mbox{for}\quad  z< 0 \nonumber\\
    &=&  0 \quad \mbox{otherwise}    \label{eq-step2}
\end{eqnarray}
\end{mathletters}

This equation was also presented in earlier work
\cite{kn:mln}  in the context of the surface dynamics
of an evolving sandpile. The two terms in the transfer term $T$ 
represent two different physical effects which we
will discuss in turn.  The first term represents
the effect of tilt, in that it models the transfer
of particles from the boundary layer at 
the `stuck' interface to the flowing
species whenever the local slope is steeper than some
threshold (in this case zero, so that negative slopes
are penalised). The second term is restorative in its
effect, in that in the presence of `dips' in the interface
(regions where the slope is shallower, $i.e.$ more positive than
the zero threshold used in these equations), the flowing
grains have a chance to resettle on the surface and replenish
the boundary layer \cite{kn:ambook}.
We notice that because one of the terms in $T$
is independent of $\rho$ we are no
longer restricted to a coupling which
exists only in the presence of flowing grains:
$i.e.$ this model is applicable to intermittent
avalanches when $\rho$ may or may not always
exist on the surface. 
In the following we examine the effect of
this interaction on the profiles of
$h$ and $\rho$ respectively.

The complexity of the transfer term
with its discontinuous functions precludes
any attempts to solve this model along
the lines of the earlier ones. We make some
remarks here, however, on the likely
critical behaviour of this model.

We observe that the transfer term
$$ T= -\lambda\rho(\na h)_{+} - \nu(\na h)_{-} $$
can be thought of as a formal infinite series
by invoking a suitable representation for
the Heaviside step functions in Eq.(23).
We are then led to consider 
the following more general structure for
the transfer term $T$,
\be
T = -\lambda \rho (\na h) - \nu (\na h)-\sum _{n=1}^{\iff}\nu_{n}
(\na h)^{n+1} -\rho\sum_{n=1}^{\iff}\lambda_n(\na h)^{n+1}
\label{eq-trans}
\ee
Note however that this is not a very well-defined expansion because
the coefficients in the infinite series could well be
very large, if not infinite.
However, given this disclaimer, we can still make
the following comments in the spirit of self-consistency 
$i.e.$ subject to numerical verification. 

If $\lambda \rho(\na h) $ were the only nonlinearity, as in Case B,
we would have $ z_h=1$.
Using $ h \sim x^{\alpha_h} $ and $ \rho \sim x^{\alpha_\rho}$, we see
$\lambda\rho(\na h) $ is a more relevant nonlinearity
than $\nu_1(\na h)^2 $, the leading nonlinear term
in the expansion of $(\na h)_-$, and is likely to be 
the controlling nonlinearity
for the extreme long wavelength behaviour.
Fig.\ref{fig-lambda_C}  shows
that  the $\lambda$ vertex never
renormalises in the presence of the 
the  KPZ term $\nu_1(\na h)^2 $, so
that $z_h$ is always fixed at unity. However,
the KPZ vertex corresponding to $\nu_1(\na h)^{2}$  has distinct
behaviour in different wavevector ranges.
In the range where the vertex
renormalises, we cannot say much
about the behaviour of $\alpha_h$;
however, in the range where it does {\it not}
renormalise, 
we might imagine that 
normal KPZ hyperscaling $\alpha_h+z_h= 2$
would be restored. This,
 with $ z_h=1 $, would give $\alpha_h=1 $.

If $z_h = 1$, we can write
the scaling relation $S_h(k,\omega=0)$  for
the double Fourier transform at zero frequency as
$$ S_h(k,\omega=0) \sim k^{-2-2\alpha_h}$$
which, in the regime where the KPZ hyperscaling
holds, should look like
$ S_h(k,\omega=0)\sim k^{-4}. $

We now try to obtain additional insights into 
the behaviour of these equations
using the Hartree-Fock approximation. 
The spirit of Hartree-Fock is to replace 
nonlinear terms by linear ones with coefficients 
that are generally determined 
self-consistently. To undertake that here, 
we note that the step functions (Eq.\ref{eq-trans}) 
give rise to nonlinearities and hence the 
simplest thing to do is to replace them 
by an expectation value 
(the argument of the  step function is a 
random variable and hence this is
an acceptable approximation). We represent this 
expectation value by 
a number $c$ with $0 < c < 1$. The equations 
of motion thus read
\begin{mathletters}
\begin{eqnarray}
{\pa h\over \pa t} &=& D_h \na^2 h -\lambda^\pr\rho\na h -\nu^\pr \na h +\eta_h(x,t) \label{eq-bcre1}\\
{\pa \rho \over \pa t} &=& D_\rho \na^2 \rho + \lambda^\pr\rho\na h + \nu^\pr\na h 
\label{eq-bcre2}
\end{eqnarray}
\end{mathletters}
with $\lambda^\pr = c \lambda$ and $\nu^\pr = (1-c)\nu$  and are identical to  the ones 
studied by Bouchaud {\it et al.} \cite{kn:bcre}.
We expect at least in some regime of Eqs.(21) to 
reproduce the mean-field 
results appropriate to Eqs.(\ref{eq-bcre1},\ref{eq-bcre2}). 

\subsubsection{ Results for the single Fourier transforms}

The single Fourier transforms $S_h(k,t=0)$ (Fig.\ref{fig-C1}) 
and $S_h(x=0,\omega)$ 
(Fig.\ref{fig-C2}) show power-law behaviour corresponding to
\[   S_h(k,t=0)  \sim  k^{-2.56\pm 0.060}   \]
\[   S_h(x=0,\omega)  \sim  \omega^{-1.68\pm 0.011} \]
which implies that the roughness and the growth exponents are 
given by respectively $\alpha_h=0.78\pm 0.030$ and $\beta_h=0.34\pm 0.005$.
This suggests $z_h = \alpha_h/\beta_h \approx 2 $ contradicting the prediction 
of $z_h = 1$ by perturbative methods 
and suggesting that the mean-field 
approach outlined in the above might be more appropriate.
We discuss this further in what follows.

However, the small $k$ limit of $S_h(k,t=0)$ indicates a downward curvature
and thus a deviation from the linear behaviour at  
higher $k$ (Fig.\ref{fig-C1}). This curvature, which had also been
observed in  previous work \cite{kn:mln}, indicates a smaller
roughness exponent $\alpha_h$ there, $i.e.$  an asymptotic {\it smoothing}.

\subsubsection{ Results for the double Fourier transforms}

The double Fourier transforms $S_h(k,\omega=0)$ (Fig.\ref{fig-C3}) 
and $S_h(k=0,\omega)$ (Fig.\ref{fig-C4}) show
power-law behaviour corresponding to
\begin{eqnarray}
 S_h(k=0,\omega) & \sim  {\omega^{-1.80\pm 0.007}}& \nonumber \\ 
 S_h(k,\omega=0) & \sim k^{-4.54\pm 0.081} & \quad 
     \mbox{for large wavevectors} \nonumber \\
        & \sim \mbox{constant} & \quad 
     \mbox{for  small wavevectors} \nonumber 
\end{eqnarray}

The double Fourier transform $S_h(k=0,\omega)$ shows 
the usual $\omega^{-2}$
behaviour that we have seen before in 
Eqs.(\ref{eq-skw_A}) and (\ref{eq-hw2}) which we have already 
discussed earlier.

The structure factor $S_h(k,\omega=0)$ signals a  dramatic behaviour
of the roughening exponent $\alpha_h$, which crosses over from  
\begin{itemize}
\item   A value of  $1.3 $ indicating anomalously 
large roughening at intermediate wavevectors, to 
\item  A value of about $-1$ for small wavevectors indicating 
asymptotic hypersmoothing.
\end{itemize}

The anomalous roughening $\alpha_h=1$  seen here 
is consistent with that observed via the
single Fourier transform (Fig.\ref{fig-C1})
and suggests, via the perturbative arguments
given previously, that $z_h=1$.
However, if we assume $z_h = 2 $ according to
the results of the single Fourier transforms given above,
this would lead to an $\alpha_h$ of about 0.8,
in agreement with the values obtained both 
via single Fourier transforms in the present paper,
and in \cite{kn:mln}.
In either case, our values of $\alpha_h$ (either 1.3 or 0.8) suggest
anomalous roughening of the interface at moderately large
wavevectors.
 
The anomalous smoothing obtained here ($\alpha_h \sim -1$ if $z_h \sim 1$, and
$\alpha_h \sim -1.5$ in the event that $z_h$ is taken to be 2) is also consistent with 
the downward curvature in the single Fourier transform $S_h(k,t=0)$, as both 
imply a negative $\alpha_h$; we mention also that 
the wavevector regime where this smoothing 
is manifested is almost identical in both 
Figs.\ref{fig-C1} and \ref{fig-C3}.

Since we expect that the anomalous smoothing 
results from a failure of the 
expansion of the step functions along the lines
of Eq.(\ref{eq-trans}), this underlines our expectation that  
the mean-field solution of Eqs.(\ref{eq-bcre1},\ref{eq-bcre2}) 
would  capture at least some of the flavour of this regime.
We have therefore solved the mean-field 
equations (Eqs.\ref{eq-bcre1},\ref{eq-bcre2})
numerically, and from Fig.\ref{fig-bcre1} and Fig.\ref{fig-bcre2}
we find that there is a crossover in $S_h(k,t=0)$(Fig.\ref{fig-bcre1})
from a diffusive behaviour ($z_h=2$) at high wavevectors
to a smoothing behaviour at low wavevectors.

This behaviour is reflected in our results for Case C. 
At low frequencies the region of anomalous
smoothing  can be understood by comparison 
with the corresponding region in the mean-field equations
Eqs.(\ref{eq-bcre1},\ref{eq-bcre2}) which also manifest this. At large $k$,
$S_h(k,t=0)$ and $S_h(k,\omega=0)$ indicate anomalous 
roughening with $\alpha_h\approx z_h\approx 1 $ 
which is consistent with the 
infrared divergence discussed in the previous 
section. However, as in Case A, $S_h(x=0,\omega)$ is dominated by 
the  diffusive $z_h = 2$ 
arising from the presence of $\delta(\omega - \nu^\pr k)$ 
in the mean-field solution of Case C. 
This behaviour is corroborated by an evaluation of the 
full structure factor $S(k_i,\omega) $ (Fig.\ref{fig-C5})
which shows a distinct peak at an $\omega_i$ given 
by $ \omega_i = \nu^\pr  k_i $; this is
reminiscent of the Lorentzian obtained in Case A (Fig.\ref{fig-A1}). 
In fact, to leading order, 
 $S_h(k,\omega)$ can be fitted to a Lorentzian;
however, as we reduce the relative strength of $\nu (\na h)_{-} $
with respect to $\lambda \rho(\na h)_{+} $
the Lorentzian peaks disappear, and
we begin to see a `shoulder' 
reminiscent, as it should be, of
the behaviour observed in Case B (Fig.\ref{fig-B3}).
 This suggests that the present model
is an integrated version of the earlier
two, reducing to their behaviour in different
wavevector regimes; we speculate therefore
that there are {\it two} dynamical exponents 
($z_h = 1$ and $ z_h = 2$) in the problem. 

\section{ DISCUSSION AND CONCLUSIONS}

We have presented in the above
a discussion of three models of sandpiles, all of which
manifest asymptotic smoothing:
Cases A and C manifest this in
the species $h$ of stuck grains,
while Case B manifests this in the species $\rho$
of flowing grains.
We reiterate that the fundamental physical
reason for this is the following:
Cases A and C both contain
couplings which are independent
of the density $\rho$ of flowing
grains, and are thus applicable
for instance to the dynamical regime
of intermittent avalanching
in sandpiles, when grains occasionally
but not always flow across the `bare' surface.
In Case B, by contrast, the equations
are coupled only when there is 
continuous avalanching, $i.e.$ 
in the presence of a finite
density $\rho$ of flowing grains.

The analysis of Case A is straightforward,
and was undertaken really only to
explain features of the more complex Case C;
that of Case B shows satisfactory agreement
between perturbative analysis and simulations. Anomalies
persist however when such a comparison
is made in Case C, because the discontinuous
nature of the transfer term makes it analytically
intractable. These are removed when the analysis includes
a mean-field solution which is 
able to reproduce 
the asymptotic smoothing observed. 

We suggest therefore an experiment where
the critical roughening exponents of
a sandpile surface are measured in
\begin{enumerate}
\item  {a rapidly
rotated cylinder, in which the time between
avalanches is much less than the avalanche duration.
Our results predict that for small system sizes
we will see only diffusive smoothing, but that
for large enough systems, we will see extremely
smooth surfaces.} 
\item  {a slowly rotated cylinder where
the time between avalanches is much more
than the avalanche duration.
In this regime, the results of Case
C make a fascinating prediction:
anomalously large spatial roughening
for moderate system sizes crossing over
to an anomalously large spatial smoothing for
large systems.}
\end{enumerate}

Finally we make some speculations
in this context concerning natural phenomena.
The qualitative
behaviour of blown sand dunes \cite{kn:bag} is in 
accord with the results of
Case B, because sand moves
swiftly and virtually continuously
across their surface in the presence
of wind. By contrast,
 on the surface of a glacier,
we might expect the sluggish motion
of boulders to result in intermittent
flow across the surface, making the
results of Case C more applicable to this situation.
It would be interesting to see if the predictions
of anomalous roughening at moderate, and anomalous
smoothing at large, length scales is
applicable here.

\section*{ACKNOWLEDGEMENTS}
Parthapratim Biswas would like to 
thank the Council of Scientific and Industrial 
Research (CSIR) for financial assistance.
Arnab Majumdar acknowledges the  hospitality of SNBNCBS
during the course of this work.
Anita Mehta acknowledges the support of the National Science
Foundation, under Grant no. PHY94-07194, at the Institute
of Theoretical Physics, Santa Barbara, where part
of this research was carried out.
We thank the referee for a critical and careful
reading of the manuscript.

\appendix
\section*{}

In this appendix we discuss some of the technical points related 
with the double Fourier transform. 
We have found that the crossover that we have seen in the Eq.(5) 
would not have been observed had we been using
the single Fourier transforms
$S_h(k,t=0)$ and 
$S_h(x=0,\omega)$
for  numerical purposes. We illustrate this by writing
explicitly the  expressions for the relevant
quantities:
\begin{mathletters}
\begin{eqnarray}
S_h(k,t=0)      &\sim  & k^{-2}\label{eq-w1} \\
S_h(x=0,\omega) &\sim   & \omega^{-2} \quad \quad 
\mbox {for $\omega$ small}\label{eq-w2}\\
S_h(x=0,\omega) &\sim   & \omega^{-1.5}\quad\ \mbox {for $\omega$ large}\label{eq-w3}
\end{eqnarray}
\end{mathletters}
The examination of $S_h(k,t=0)$  (Fig.\ref{fig-A4}) on its own yields
no indication of the crossover to the smoothing
fixed point; although there is a crossover in the $S_h(x=0,\omega)$ 
graph (Fig.\ref{fig-A5}a) from $\omega^{-1.5}$ 
to $\omega^{-2}$, the analysis 
below shows that {\it both } regimes reflect diffusive behaviour,
so that the smoothing fixed point $(\alpha_h=0,\beta_h=0,z_h=1)$ is 
entirely suppressed.

The single Fourier transform $S_h(x,\omega )$ is defined by 
\[
S_h(x=0,\omega) = \int_{-\iff}^{\iff} \frac{dk}{2\pi} S_h(k,\omega) 
 = \int_{-\iff}^{\iff} \frac{dk}{2\pi} \frac{1}{D_h k^{z_h}}
\left[ \frac{D_h k^{z_h}}{(\omega - ck)^2 + D_h^2 k^{2z_h}} \right] 
\]
In the limit $\omega \rightarrow ck$ the term in the square brackets behaves like a 
$\delta$-function and thus
\begin{equation}
S_h(x=0,\omega) =  \int_{-\iff}^{\iff} \frac{dk}{2\pi}
{1\over D_h k^{z_h}} \delta(\omega-ck) 
\approx   \frac{1}{\omega^{z_h}}
\label{eq-delta}
\end{equation}
This is the origin of  ballistic behaviour in the flow term
and is responsible for two anomalies.

\begin{enumerate}

\item  Firstly, we notice from the above
that the $\delta$-function causes $S_h(x=0,\omega)$ to behave like 
$\omega^{-z_h}$.
Comparing with Eq.(\ref{eq-w2}) this leads to
$z_h = 2$. However a simple-minded
application of Eqs.(\ref{eq-1d})
would have led to the {\it wrong}
conclusion of $\beta_h = 0.5$.
Even if the correct scaling relation Eq.(\ref{eq-delta})
were employed, 
the ballistic nature of
the flow term picks out, misleadingly,
the {\it high} frequency (diffusive)
dynamical exponent in the {\it low}
frequency regime of $S_h(x=0,\omega)$ (Eq.(\ref{eq-w2})).
The low wavevector, low-frequency smoothing behaviour 
is thus entirely suppressed.

\item Secondly, spurious oscillations
are observed (Fig.\ref{fig-A5}(b)) in the
graph for $S_h(x=0,\omega)$ as a function of grid size.
A consideration of the form of the structure 
factor $S_h(x=0,\omega)$ makes it clear 
the crossover from small
to large $\omega$ should not involve
any imaginary quantities, and therefore strictly
speaking we should not see any oscillatory
behaviour in the structure factor in this limit.
However
the full form of the structure
factor $S_h(x,\omega)$ for finite $x$
{\it does} contain imaginary portions,
which are responsible for
the oscillations. 
The characteristic length and time scales in our
problem are given by  
$$ t_{0} = D_h/c^{2} \quad  x_{0} = D_h/c$$
Whenever grid sizes in time or space
are comparable to these characteristic
lengths, the profile fluctuates
across these intervals, which
is then aggravated by the shock
fronts associated with the flow term.
This results in
 oscillatory behaviour
arising from the {\it non-zero} intervals in $x$ 
associated with the sampling of the profile to generate the Fourier
transform, $S_h(x=0,\omega)$, which introduce a flavour
of $S_h(x,\omega)$ for {\it finite} $x$.
These become
increasingly violent as $c$ increases
because of the increased fluctuations
associated with the ballistic
flow term over the grids.
In order to avoid these oscillations,
one should choose grid sizes
$\Delta x $ and $\Delta t $ in such a way that they
are always less than
the characteristic scales in the problem, {\it i.e.},
 $$\Delta x \ll x_{0} \quad \mbox{ and } \quad \Delta t \ll t_{0} .$$ 
\end{enumerate}

In view of the above, it is necessary to use the double 
Fourier transform to obtain an unambiguous 
picture of the structure factor
and to pick out the asymptotic smoothing 
although this strategy might on first appearance
seem to be a computational overkill.  The overwhelming advantage is
that, by scanning the structure factor as a function of frequency
$\omega$ for a fixed $k$, one immediately sets 
two frequency scales $ck$ and $D_h k^{2}$, thus
making it possible to pick up the relevance of 
these scales in $S_h(k,\omega)$.

We also mention 
that our discussion is equally applicable
to the Kardar-Parisi-Zhang (KPZ) equation 
\cite{kn:kpz} with the addition of a flow term. Here too,
the use of the double Fourier transform  reveals 
the presence of the `smoothing' 
fixed point due to the flow term.

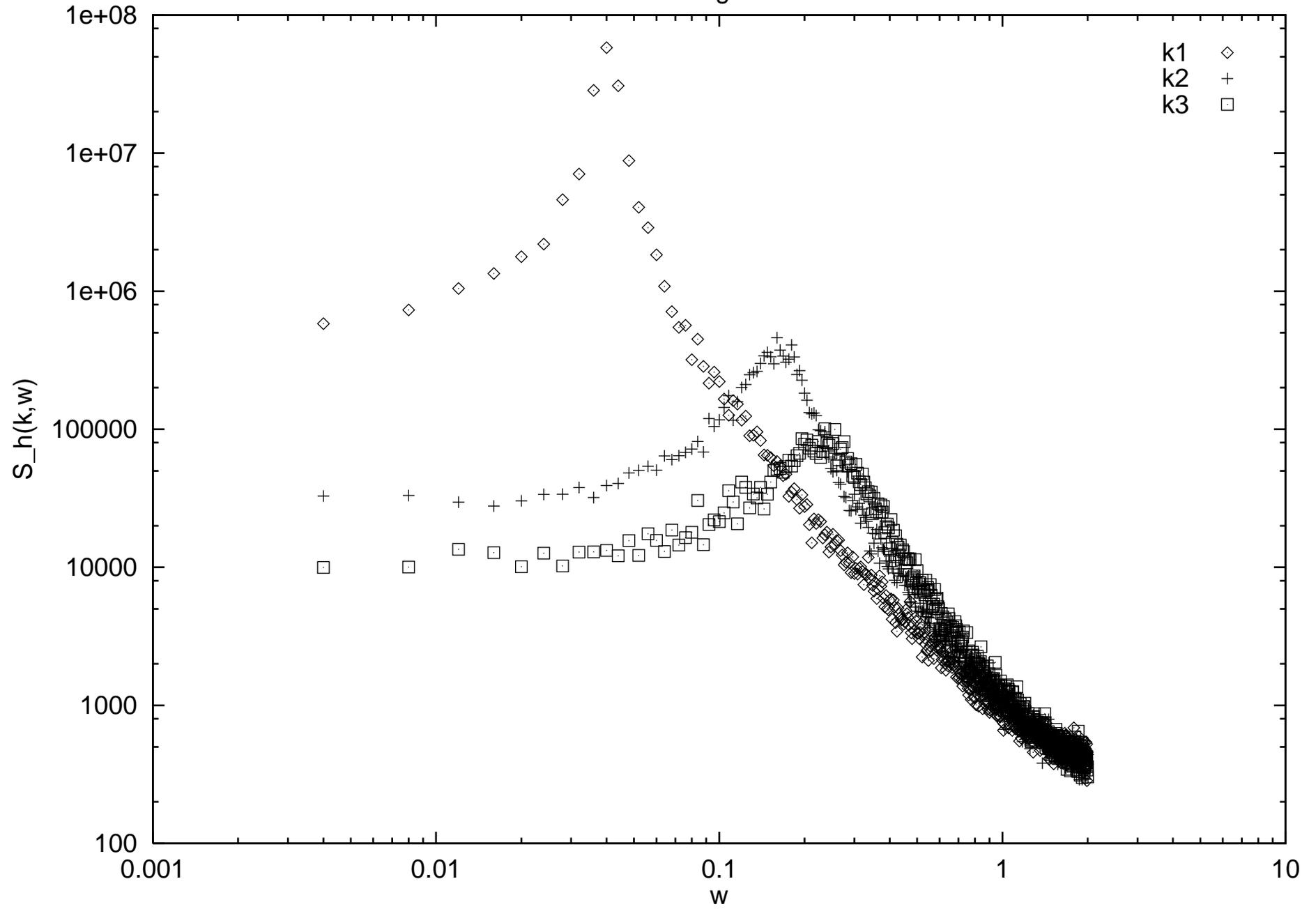
\begin{figure}
\caption{
The correlation function $S_h(k_i,\omega)$ against $\omega$
for three different wavevectors $k_1 = 0.02(\Diamond)$, $k_2 = 0.08(+)$
and $k_3 = 0.12(\Box)$ with parameters 
$c = 2.0$, $D_h = 1.0$ and $\Delta^2=1.0$. The positions of the peaks
are given by $\omega_1 = 0.04,\omega_2 = 0.16 $ and $\omega_3 = 0.24 $ as
expected from Eq.(\ref{eq-skw_A}).
}
\label{fig-A1}
\end{figure}

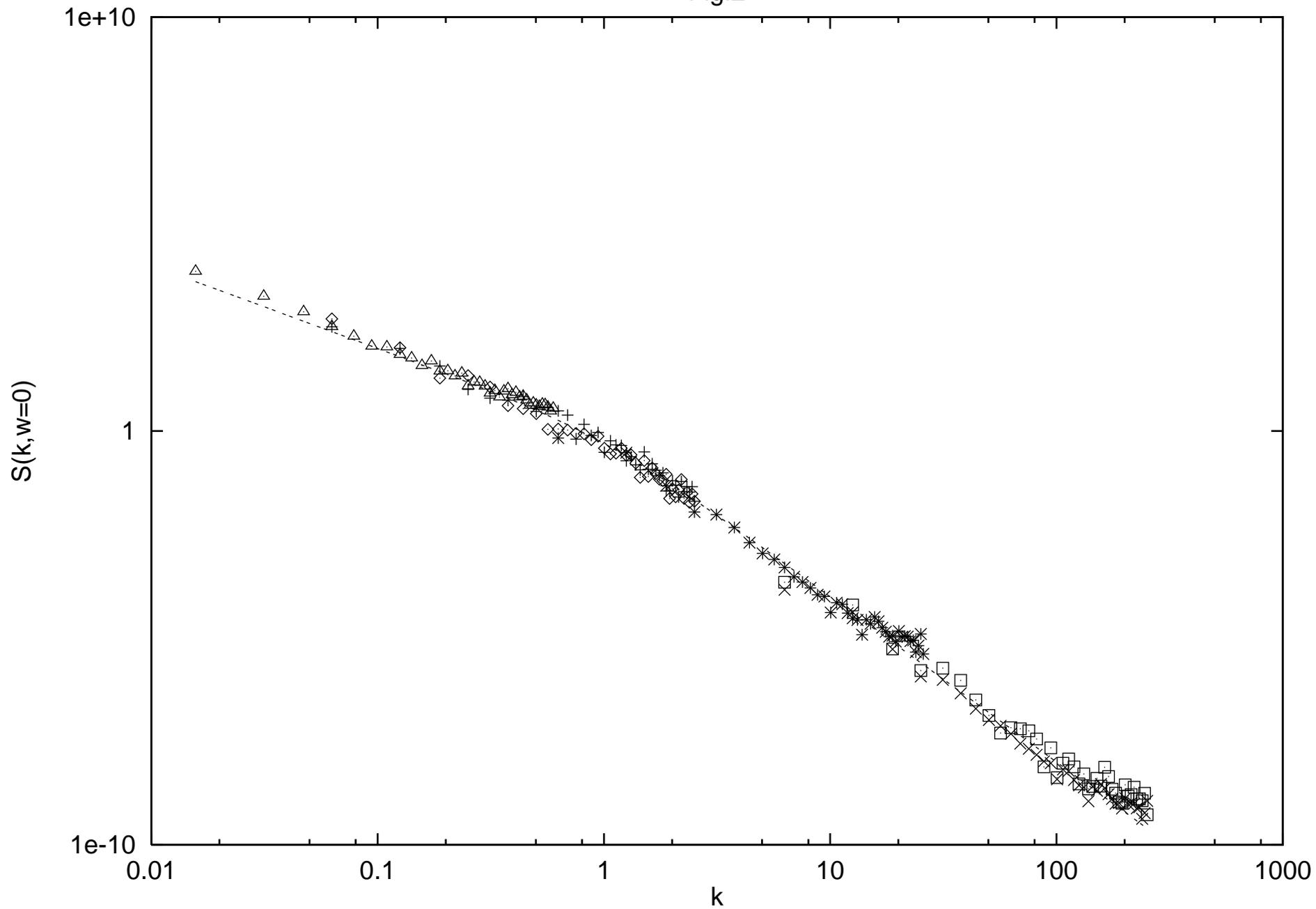
\begin{figure}
\caption{
The double Fourier transform, $S(k,\omega=0)$ obtained from Eq.(\ref{eq-few})
(Case A)
for the $h$-$h$ cor\-rel\-ation function showing the crossover from high 
to low $k$. 
The different markers in the figure correspond to  
different grid sizes $\Delta x $ to sample distinct
regions of $k$ space; thus the markers $\bigtriangleup$,
$\times$, and 
$\Box$ correspond to decreasing grid sizes and
increasing wavevector ranges.
 The parameters used in the calculation are
$c = D_h = \Delta^2 = 1.0 $ and the characteristic wavevector is
$k_0 = \frac{c}{D_h} = 1.0.$ The dashed line is a plot of $S_h(k,\omega=0)$
vs $k$ for Case A 
with appropriate parameters, to serve as a guide to the eye.
}
\label{fig-A2}
\end{figure}

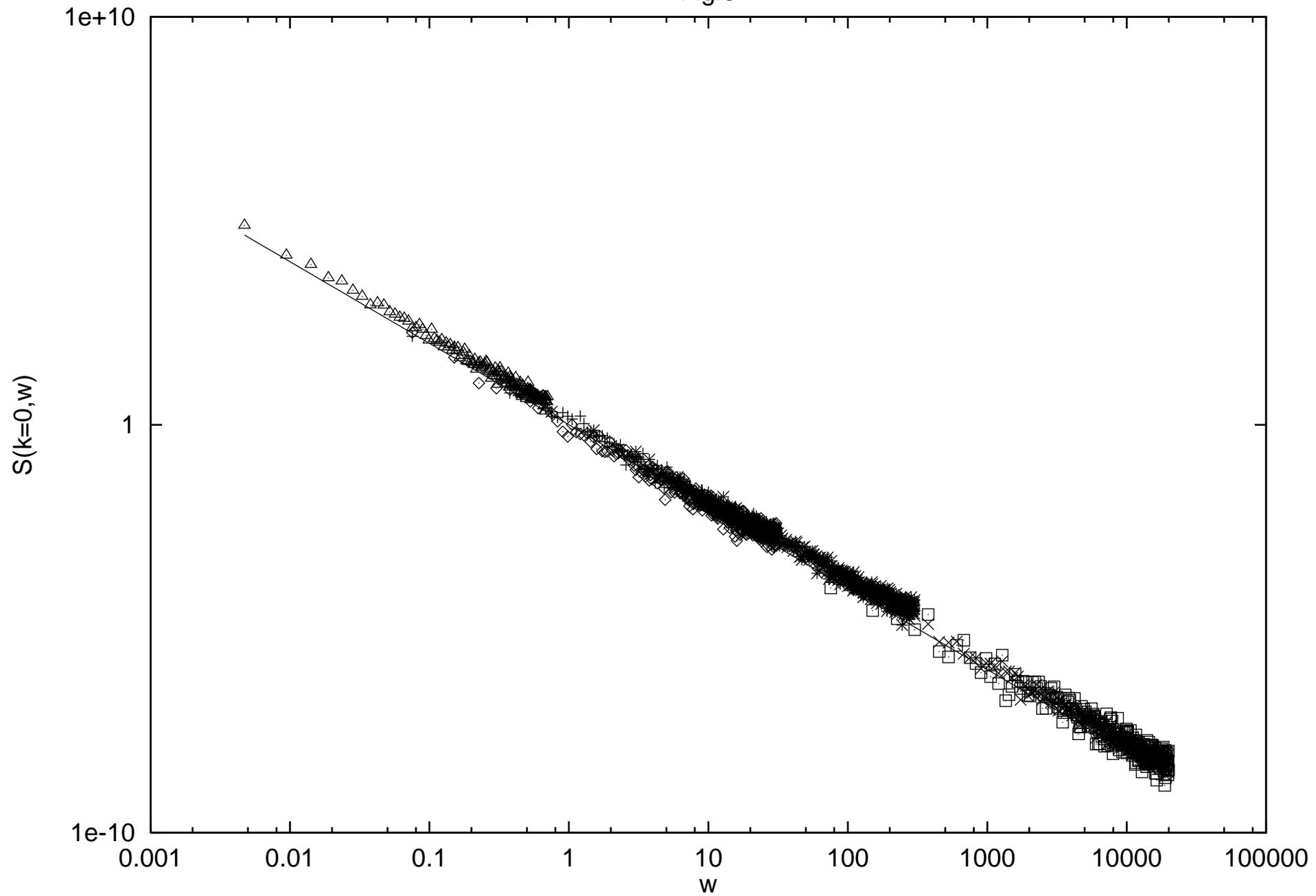
\begin{figure}
\caption{
The double Fourier transform,
$S(k=0,\omega)$ vs $\omega$  obtained from Eq.(\ref{eq-few})
(Case A)
for the $h$-$h$ correlation function. 
The different markers in the figure correspond to  
different grid sizes $\Delta t $ to sample distinct
regions of $\omega$ space; thus the markers $\bigtriangleup$,
$\times$, and 
$\Box$ correspond to decreasing grid sizes and
increasing frequency ranges.
The solid line is a plot of $S_h(k=0,\omega)$
vs $\omega$ for Case A
with appropriate parameters, to serve as a guide to the eye.
The parameters are 
$c = D_h = \Delta^2 = 1.0$.
}
\label{fig-A3}
\end{figure}

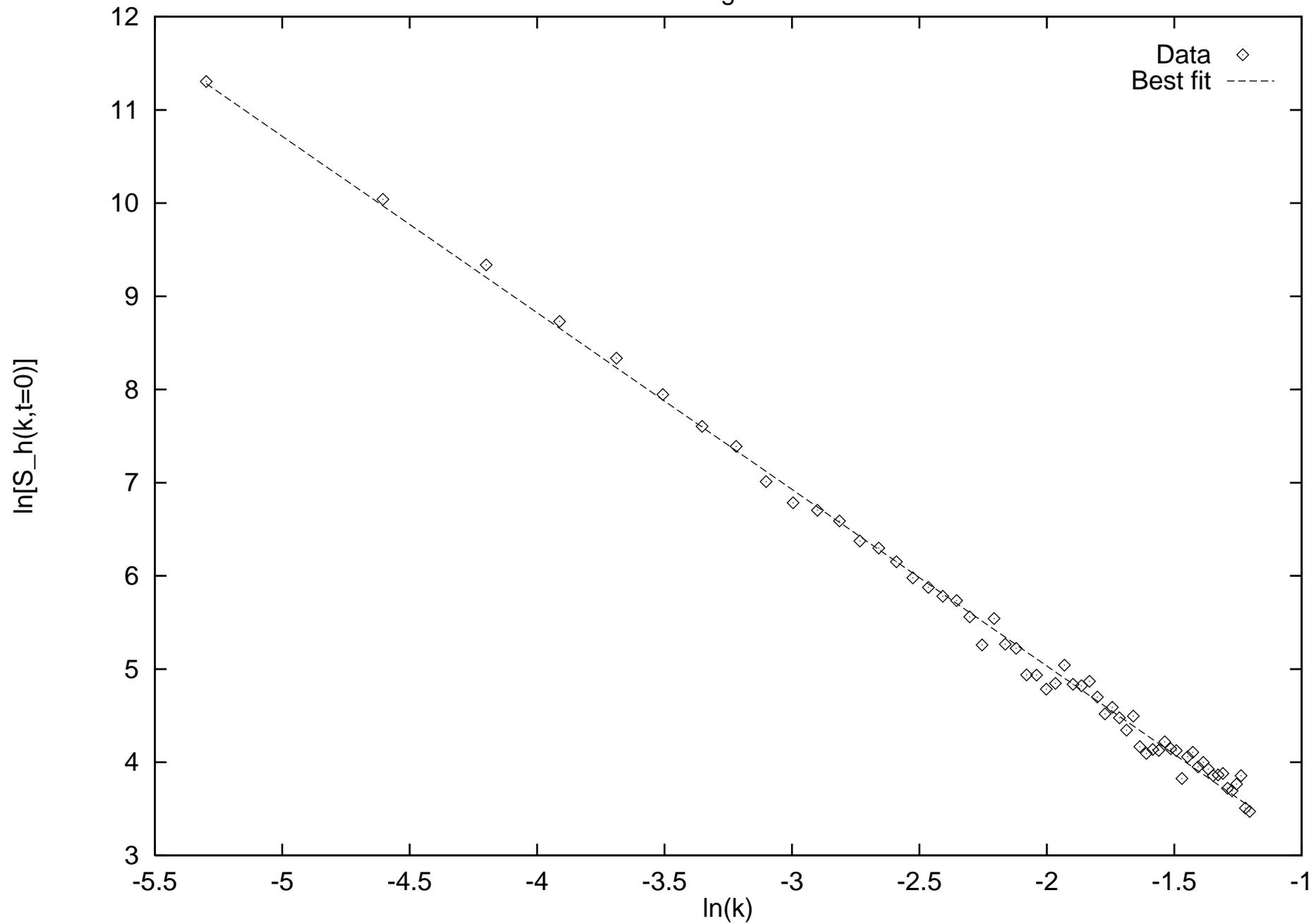
\begin{figure}
\caption{
 Log-log plot of the single Fourier transform $S_h(k,t=0)$ vs $k$ obtained from
Eq.(\ref{eq-few}) (Case A) with parameters $c = D_h = \Delta^2 = 1.0$.
The best fitted line shown in the figure is given by a 
slope  of $-1-2\alpha_h = -1.90\pm 0.016$. The characteristic wavevector 
$k_0$ is given by  $k_0 = \frac{c}{D_h} = 1.0 $
}
\label{fig-A4}
\end{figure}

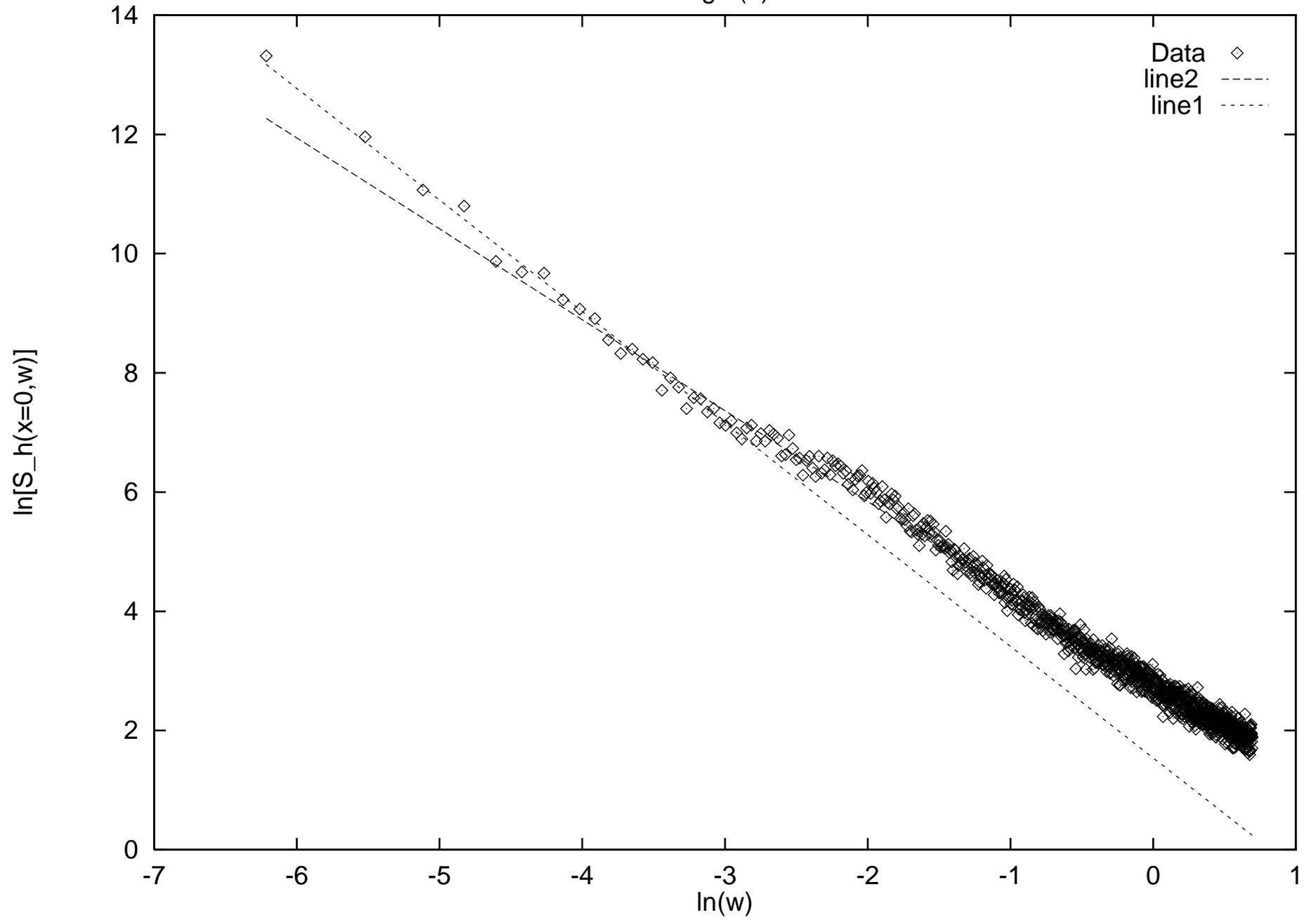
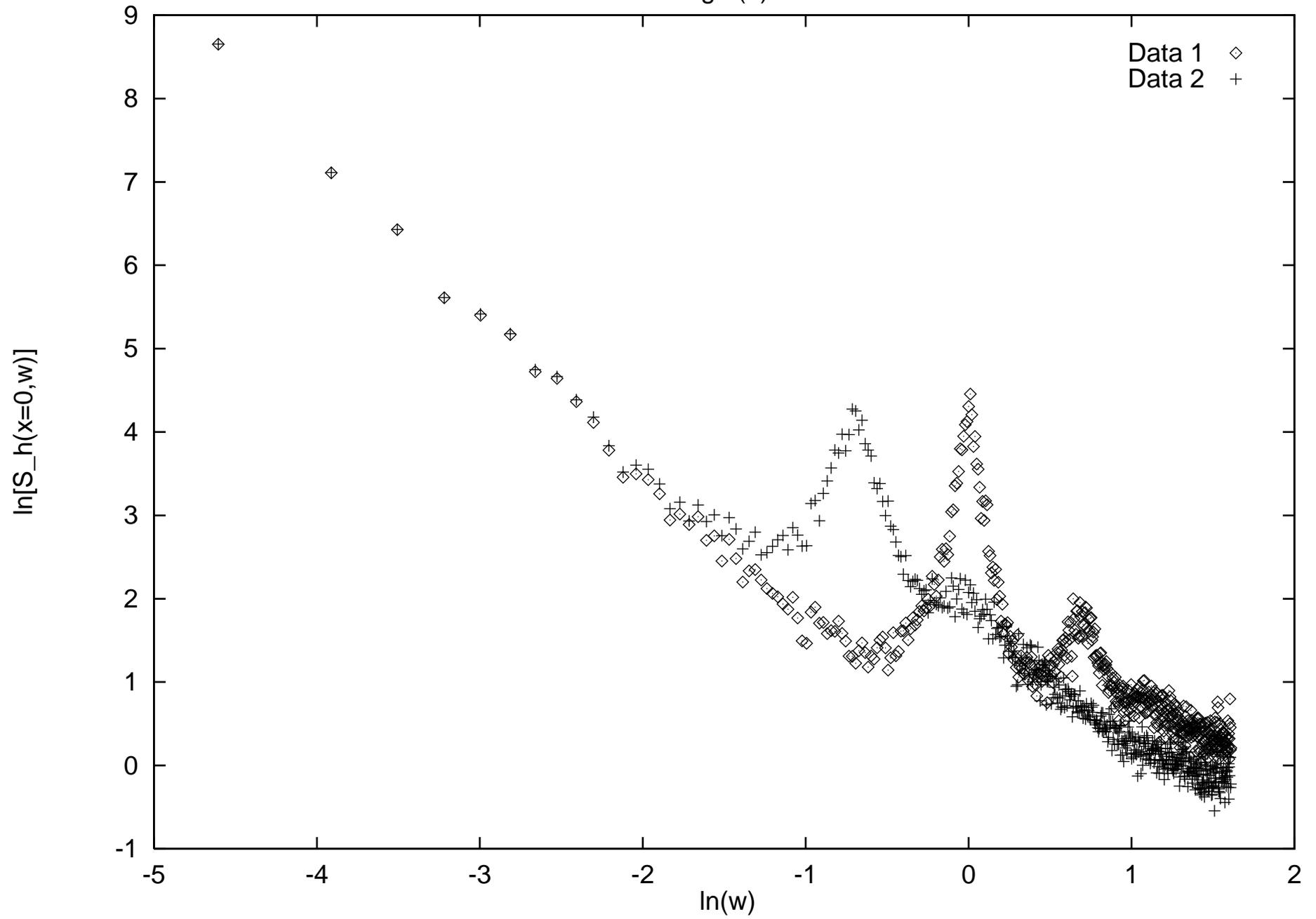
\begin{figure}
\caption{
(a)Log-log plot of the single Fourier transform $S_h(x=0,\omega)$ vs $\omega$ 
obtained from Eq.(\ref{eq-few}) showing a slow crossover. Lines
1 and 2 in the figure are the best fits in the low and high 
$\omega$ regions with slopes $-1-2\beta_h = -1.87\pm
0.003 $ and $-1-2\beta_h = -1.525\pm 0.006$ respectively.\\ \\
(b) 
Log-log plot of the single Fourier transform $S_h(x=0,\omega)$ vs $\omega$
for two different values of $c$; $c=10$ and $c=5$ for data sets 1 and 2 respectively.
Note the increase in oscillation for increasing values of $c$. The other parameters are 
$D_h = \Delta^2 = 1.0 $.
}
\label{fig-A5}
\end{figure}

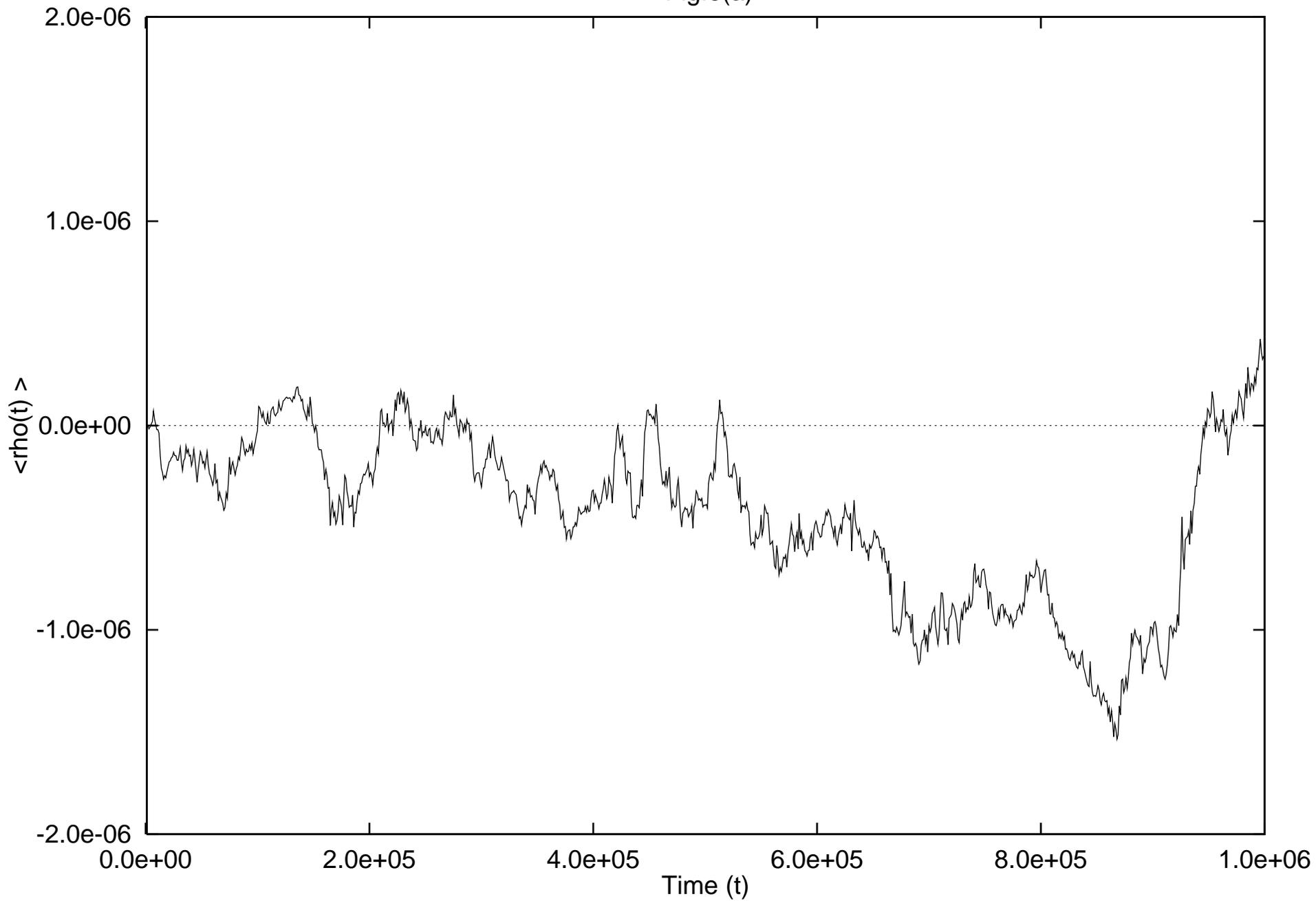
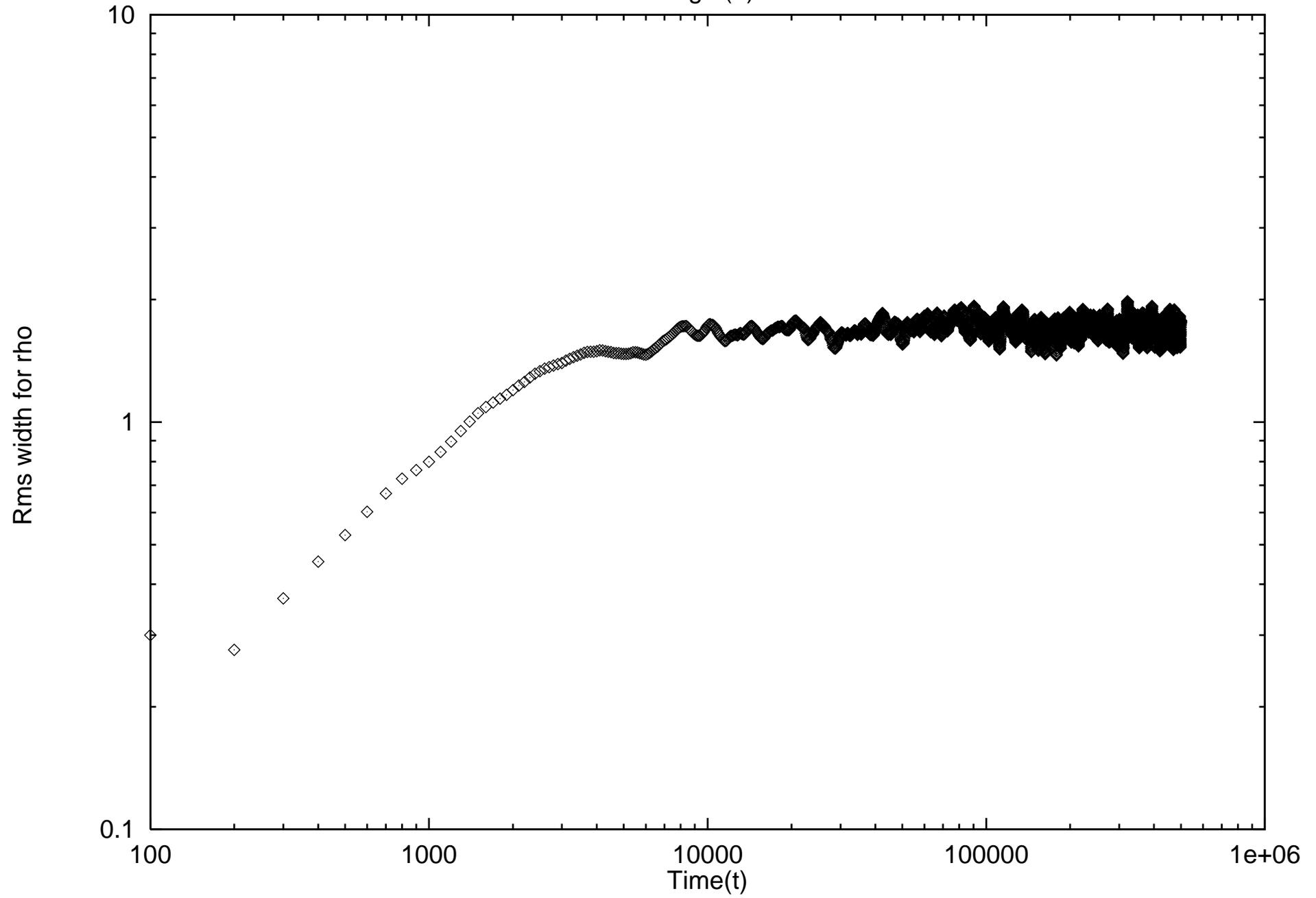
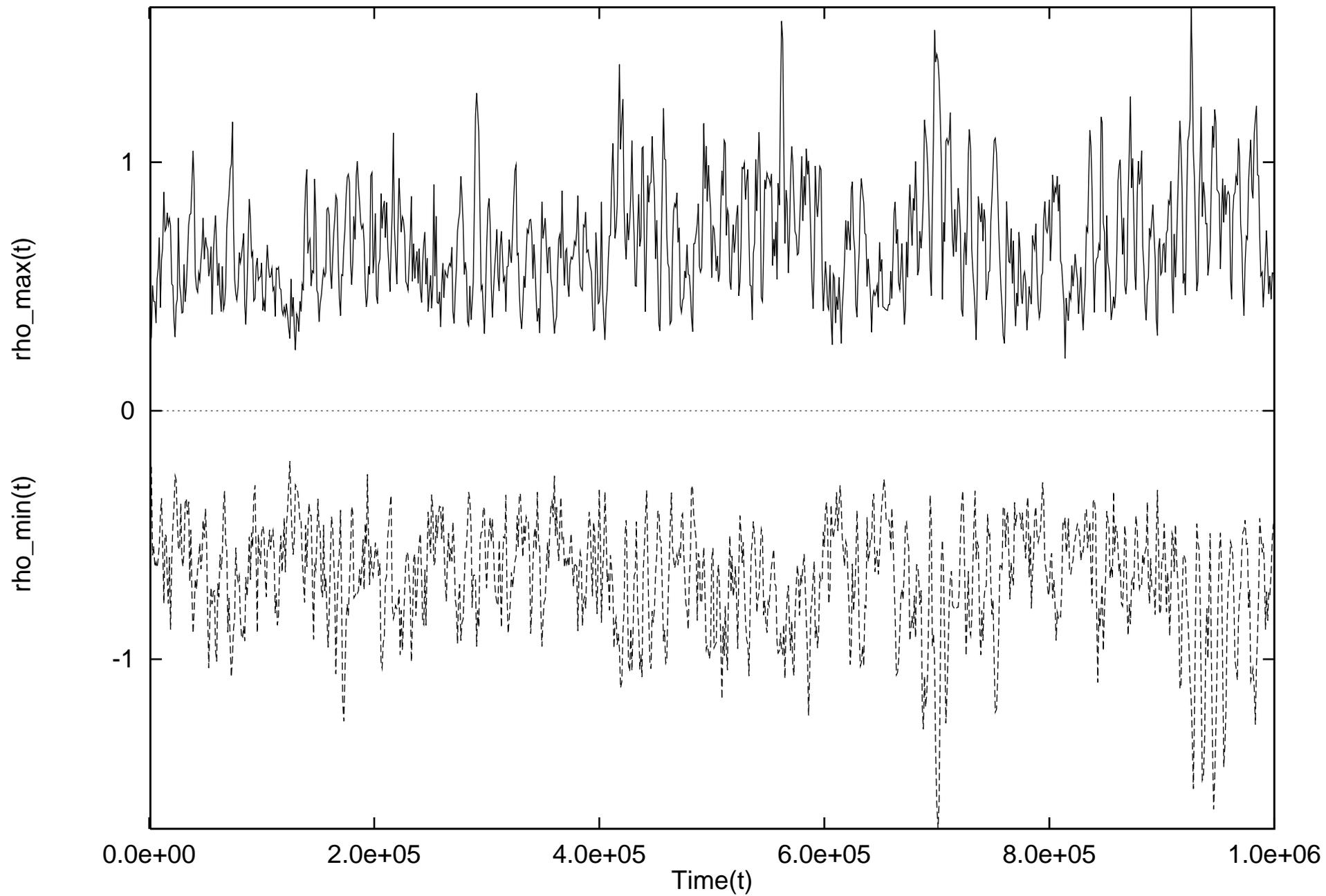
\begin{figure}
\caption{
(a) The behaviour of $<\rho(t)>$ as a function of time t.
Here $<\rho(t)>$ is the average over the sandpile surface 
of 100 sample configurations. The grid size 
$\Delta t = 0.005$  and
$c = \Delta^2 = D_h = 1.0$. \\
(b) The root mean square width $\rho_{rms}(t) = (<\rho^2>-<\rho>^2)^{\frac{1}{2}}$ 
against time t over 100 sample configurations with parameters $c = \Delta^2 = 
D_{\rho} = D_h = 1.0$ \\.
(c) The variation of $\rho_{max}(t)$ and $\rho_{min}(t)$ with time t. 
$\rho_{max}(t)$ and $\rho_{min}(t)$ are respectively the maximum and minimum values of $\rho$
for a given configuration of the sandpile at time t.
Again, $c=D_h=D_{\rho}=\Delta^2 =1.0$.
}
\label{fig-A6}
\end{figure}

\begin{figure}
\caption{One-loop diagrams for
(a) $\Sigma_\rho(k,\omega)$, the self-energy in $\rho$,
(b) $\Sigma_h(k,\omega)$, the self-energy in $h$
for the coupled equations of Case B (Eq.(6)).
(c) The glossary for the diagrams shown in Fig.7(a,b) and Fig.8. For example, the 
propagators for the $h$ and $\rho$ variables are represented by 
solid and  dashed lines respectively, with a right arrow. Additionally
there are diagrammatic definitions for the vertex and for 
the correlation functions for the $h$ and $\rho$ variables.  
}
\label{fig-fd1}
\end{figure}

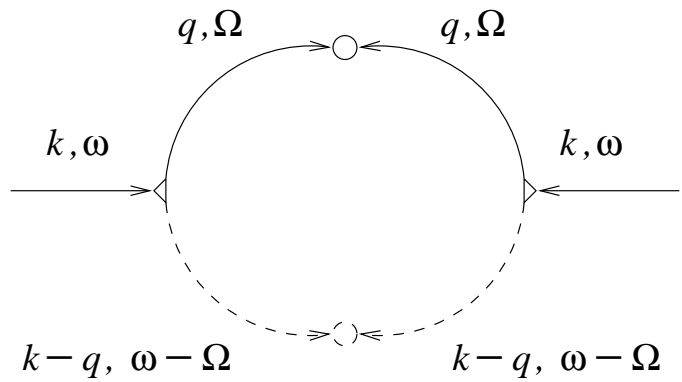
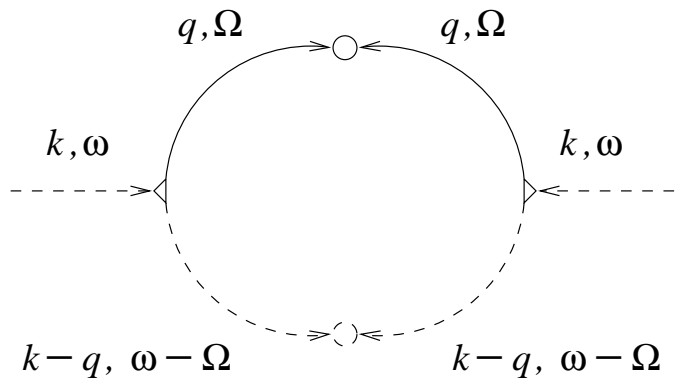
\begin{figure}
\caption{One-loop diagrams for
(a) $S_h(k,\omega)$, the $h-h$ correlation function,
(b) $S_\rho(k,\omega)$, the $\rho-\rho$ correlation function
for the coupled equations of Case B (Eq.(8)).
}
\label{fig-fd2}
\end{figure}

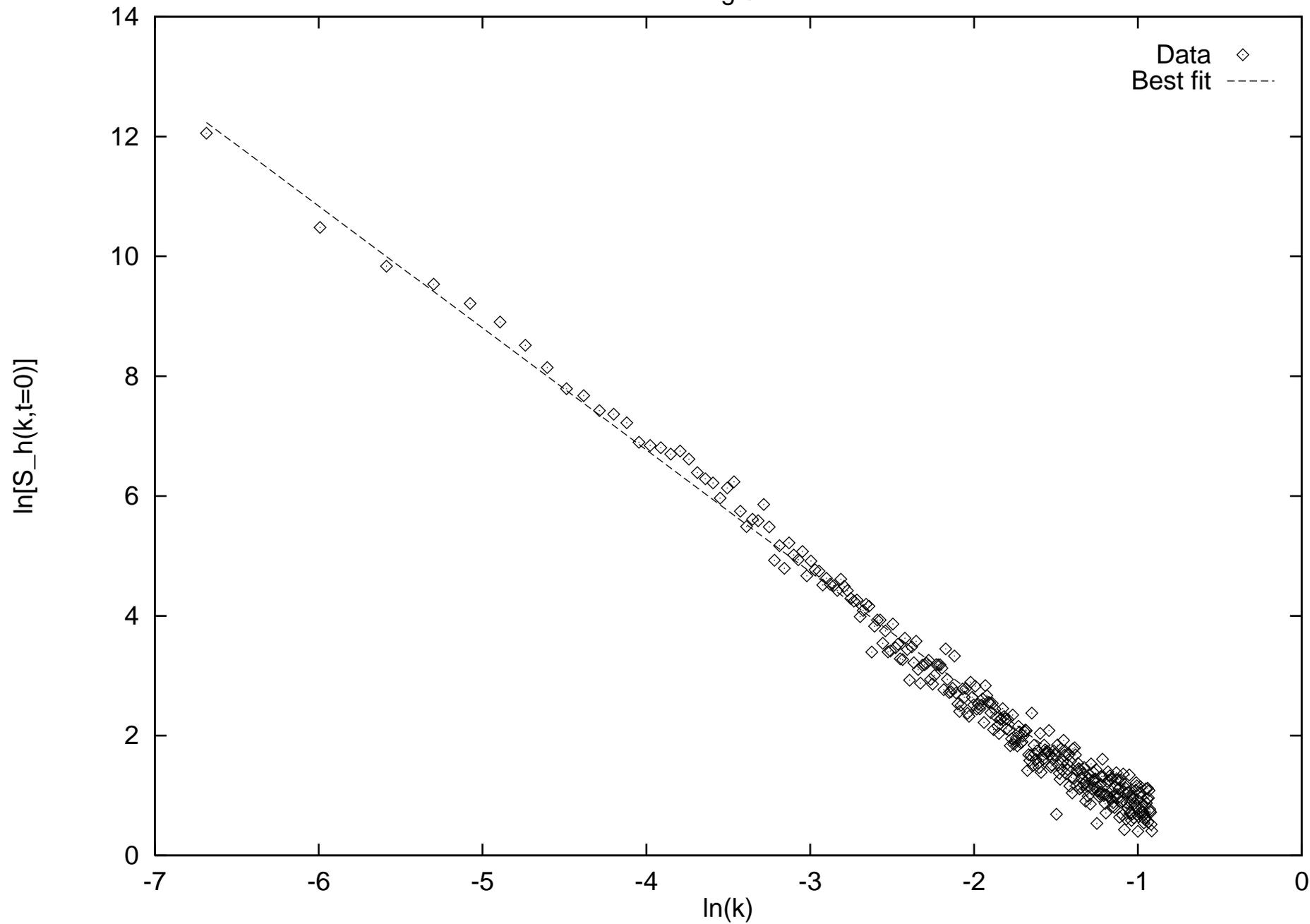
\begin{figure}
\caption{
Log-log plot of the single Fourier transform $S_h(k,t=0)$ vs $k$
obtained from Eqs.(6) ( Case B). The best fit 
has a slope of  $-1-2\alpha_h = -2.03+\pm 0.014$. Other
parameters are $\mu = D_h = D_{\rho} = \Delta_h^2 = \Delta_{\rho}^2 = 1.0$.
}
\label{fig-B1}
\end{figure}

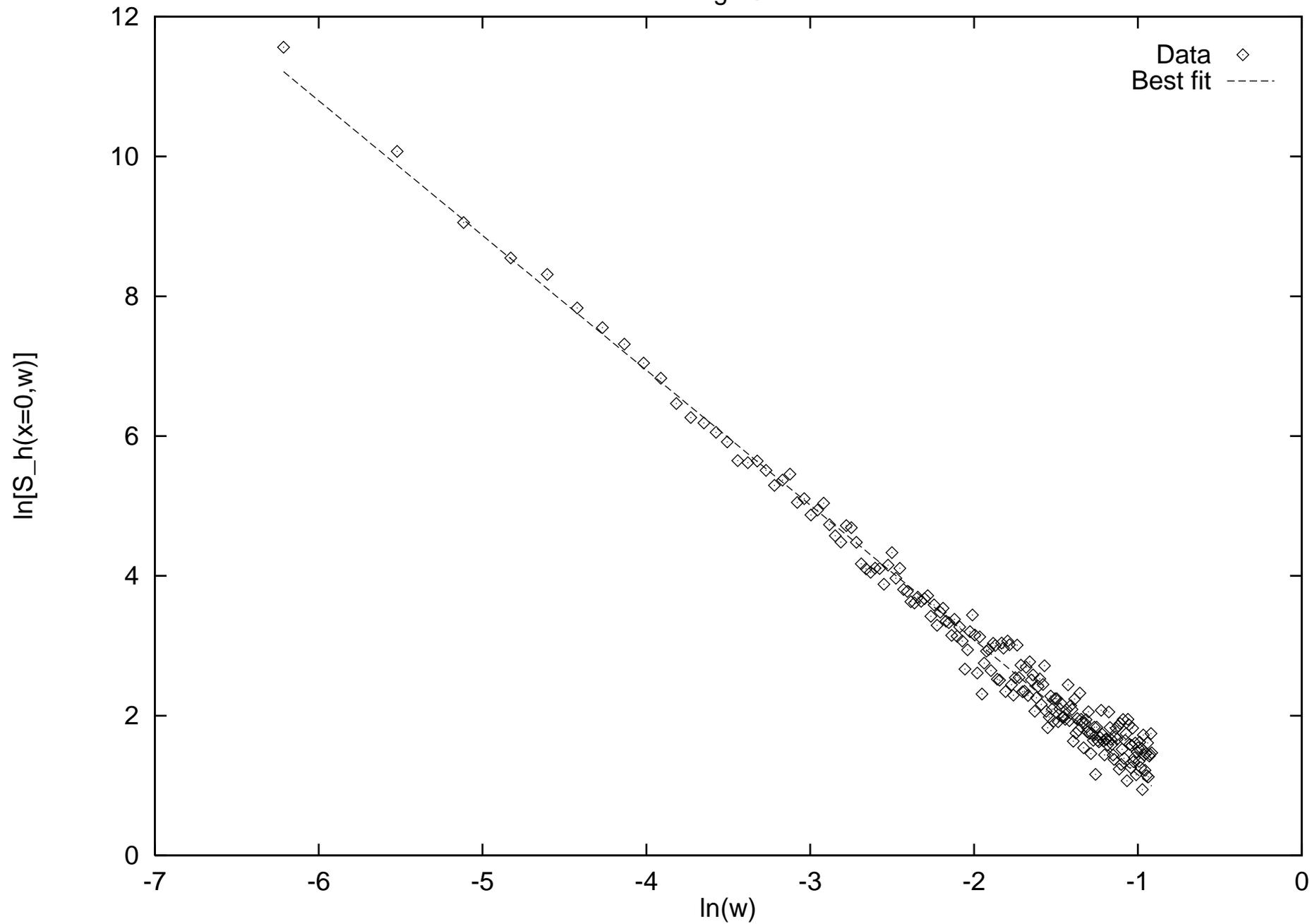
\begin{figure}
\caption{
Log-log plot of the single Fourier transform $S_h(x=0,\omega)$ vs $\omega$ for  
Case B obtained from Eqs.(6). The best fit shown in the 
figure has a  slope of $-1-2\alpha_h = 1.93\pm 0.017$.
Again $\mu = D_h = D_{\rho} = \Delta_h^2 = \Delta_{\rho}^2 = 1.0$.
}
\label{fig-B2}
\end{figure}

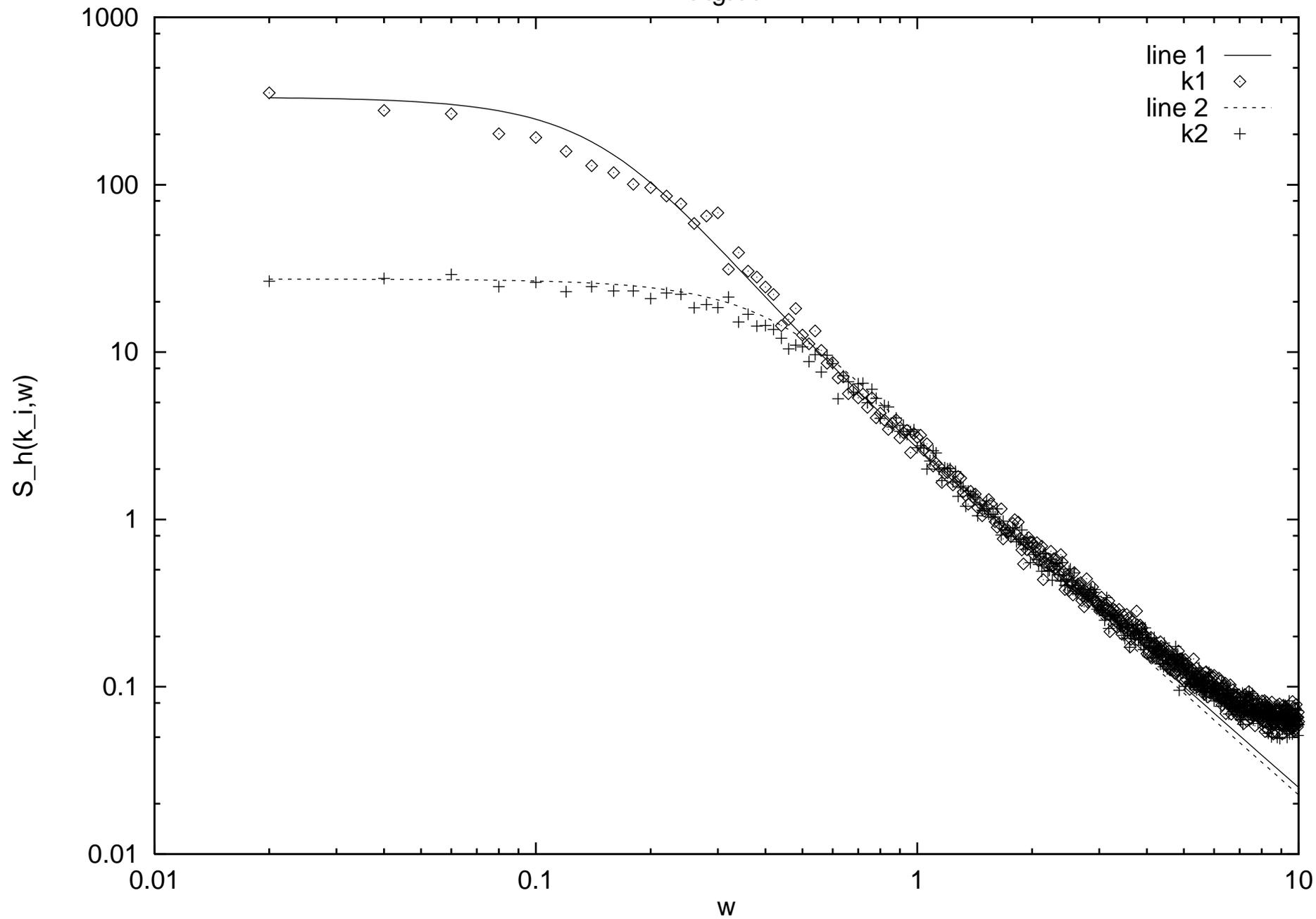
\begin{figure}
\caption{
The double Fourier transform $S_h(k_i,\omega)$ vs $\omega$  ( Case B ) 
calculated at two different wavevectors 
$k_i = 0.1 ( \Diamond ), 0.2 (+) $. The curves 
( solid(1) and dashed(2) lines ) shown 
in the figure are plots of Eq.(\ref{eq-fit}) with $\Gamma_0 = 0.4$ and 
0.5( for $k_1$ and $k_2$ respectively), to serve as a guide to the eye. 
Other parameters are $\mu = 2, \Delta_h^2 =
\Delta_{\rho}^2 = 0.1, D_h = D_{\rho} = 1.0$. 
}
\label{fig-B3}
\end{figure}

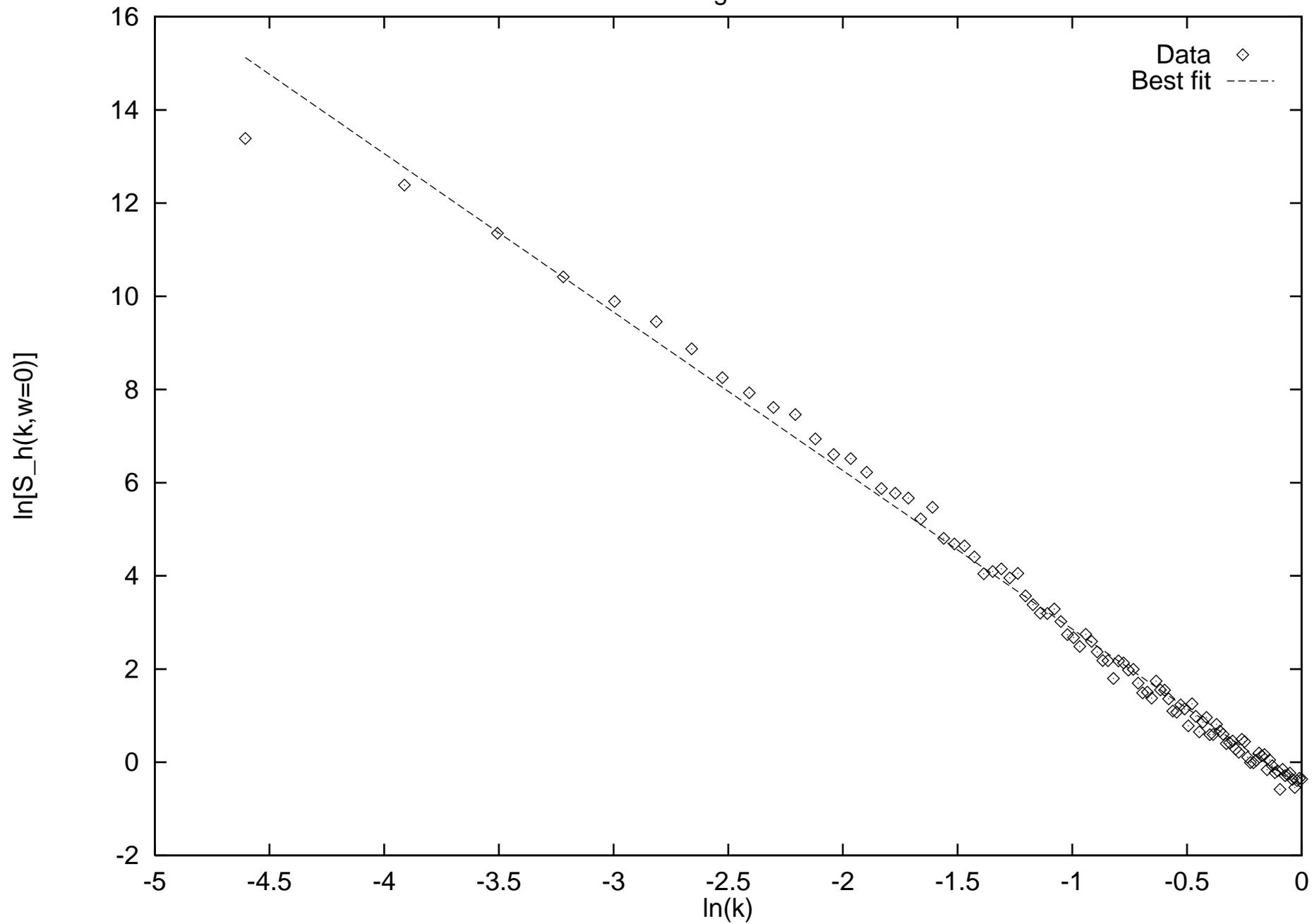
\begin{figure}
\caption{
Log-log plot of the double Fourier transform $S_h(k,\omega=0)$ vs $k$
( Case B ) obtained from Eqs.(6). The best fit has a slope of 
 $-(1+2\alpha_h+z_h) = -3.40\pm 0.029$. Again,  
$\mu = 1.0, D_h = D_{\rho} = 1.0, \Delta_h^2 = \Delta_{\rho}^2 = 0.5$.
}
\label{fig-B4}
\end{figure}

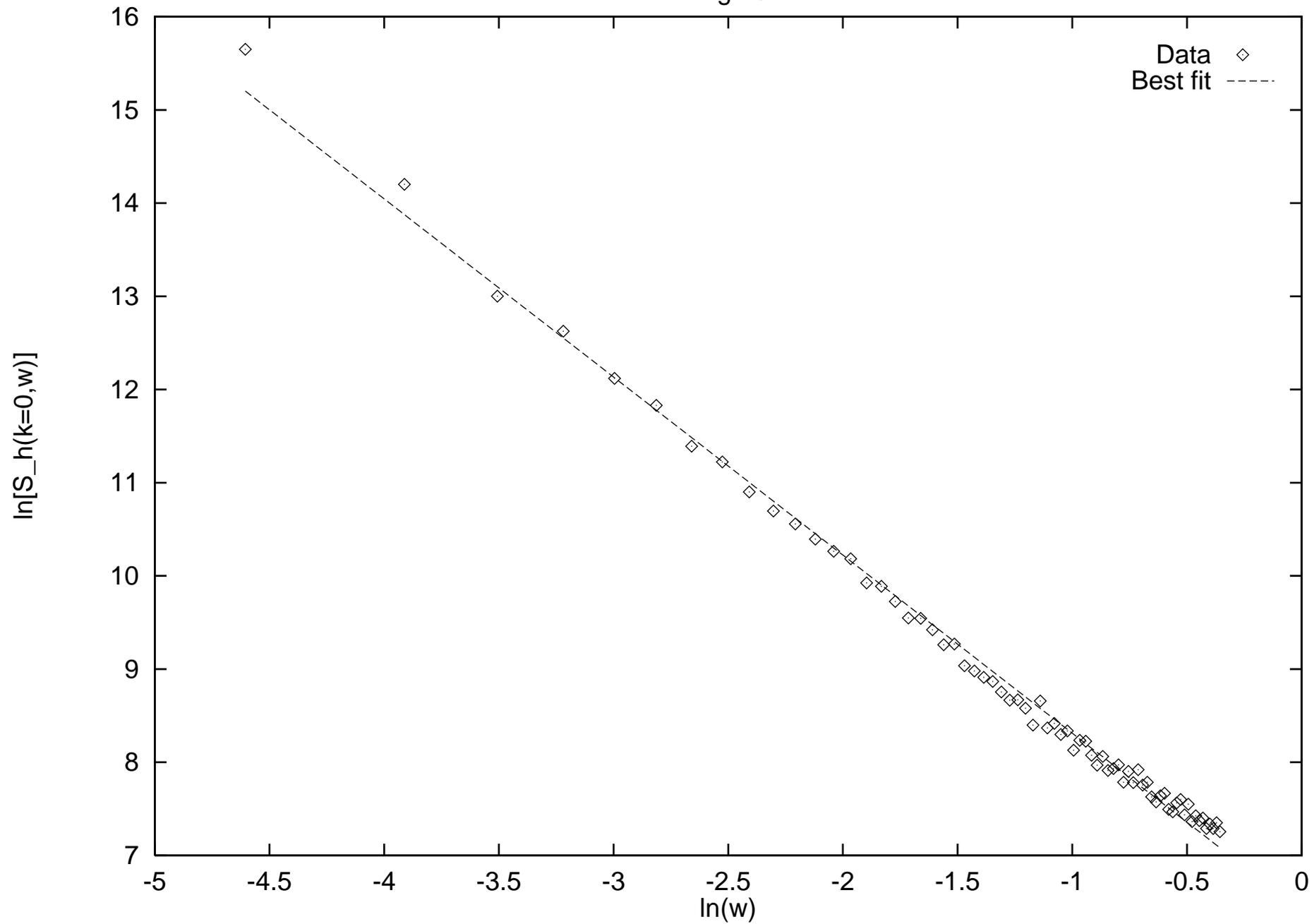
\begin{figure}
\caption{
Log-log plot of the double Fourier transform $S_h(k=0,\omega)$ vs $\omega$ 
obtained from Eqs.(6) ( Case B ). The best fit displayed in the figure has   
a slope of $-(1+2\beta_h+\frac{1}{z_h}) = -1.91\pm 0.017$.
Other parameters are  $\mu=1.0$, $D_h = D_{\rho} = 1.0$ , $\Delta_h^2 = 
\Delta_{\rho}^2 = 0.5$.
}
\label{fig-B5}
\end{figure}

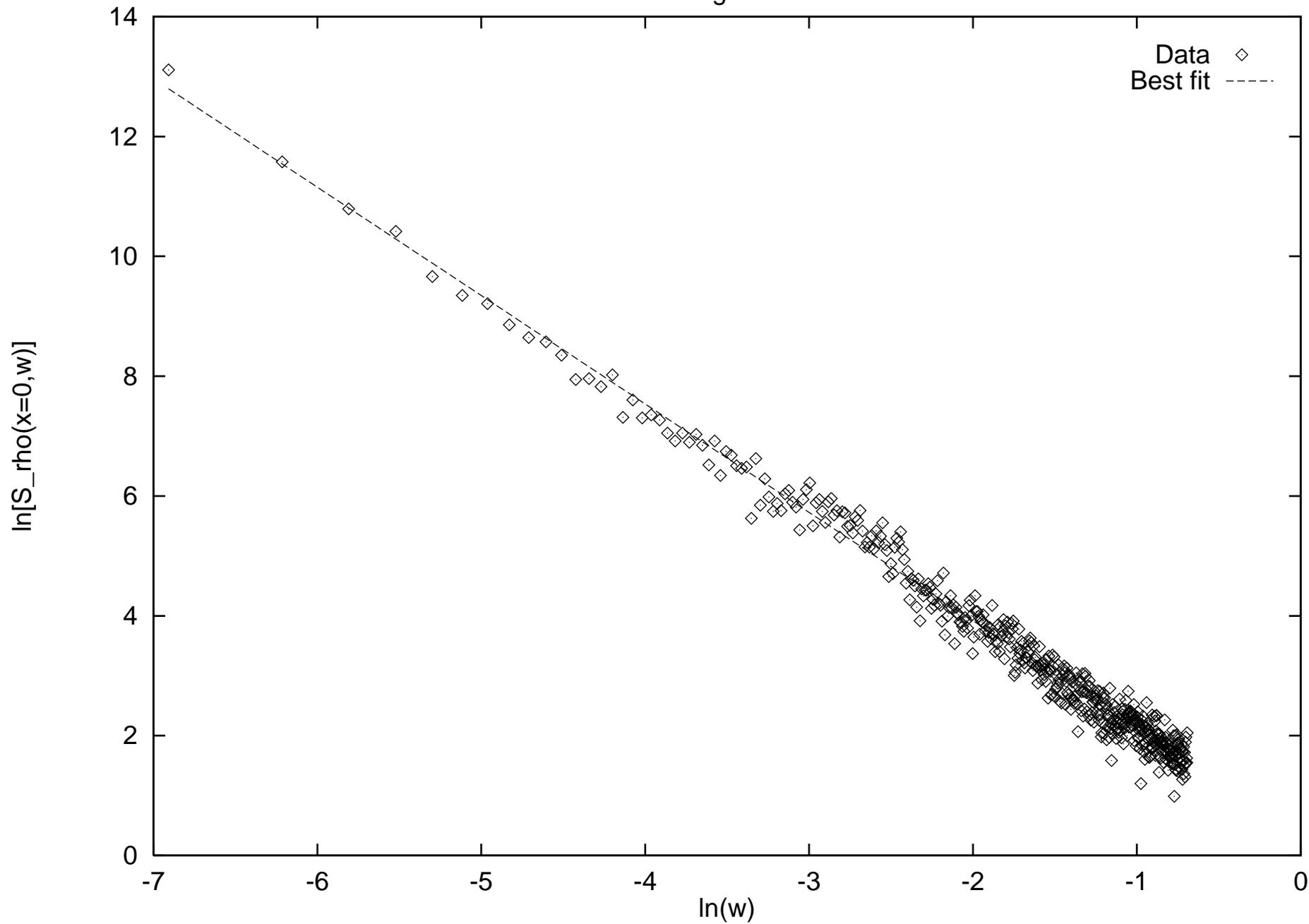
\begin{figure}
\caption{
Log-log plot of the single Fourier transform $S_{\rho}(x=0,\omega)$ vs $\omega$
obtained from Eqs.(6) ( Case B ). The best fit has a  slope of 
$-1-2\beta_{\rho} = - 1.81\pm 0.017$. Again,  $\mu=1.0, D_h = D_{\rho} = 1.0
, \Delta_h^2 = \Delta_{\rho}^2 = 0.5.$
}
\label{fig-B6}
\end{figure}

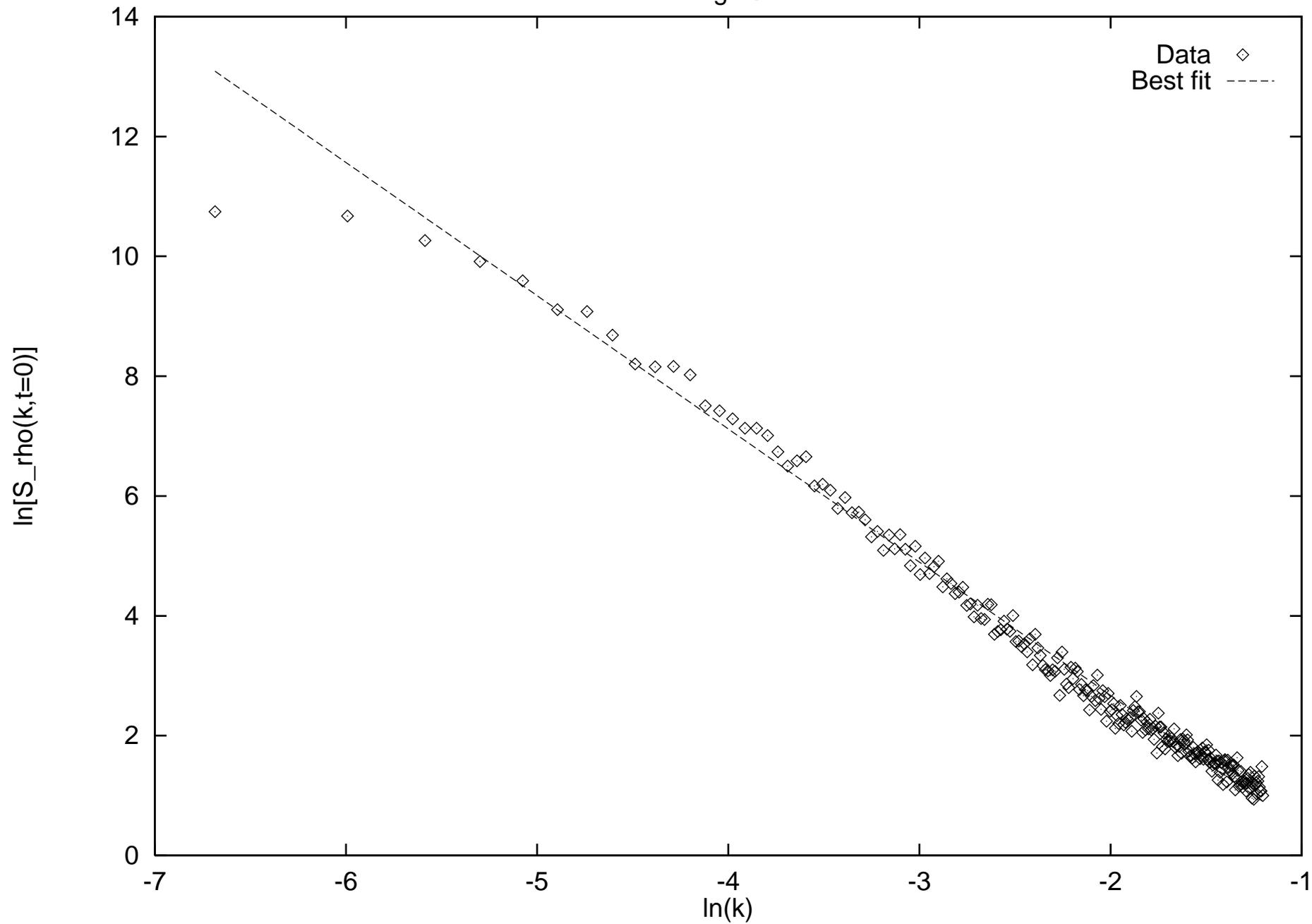
\begin{figure}
\caption{
Log-log plot of the single Fourier transform $S_{\rho}(k,t=0)$ vs $k$
( Case B ) showing a crossover from a slope of  $-1-2\alpha_{\rho} = 0$  at small 
$k$ to $-2.12\pm 0.017$ at large $k$.  
Other parameters are $\mu=1.0, D_h = D_{\rho} = 1.0, \Delta_h^2 = \Delta_{\rho}^2 = 0.5.$
} 
\label{fig-B7}
\end{figure}

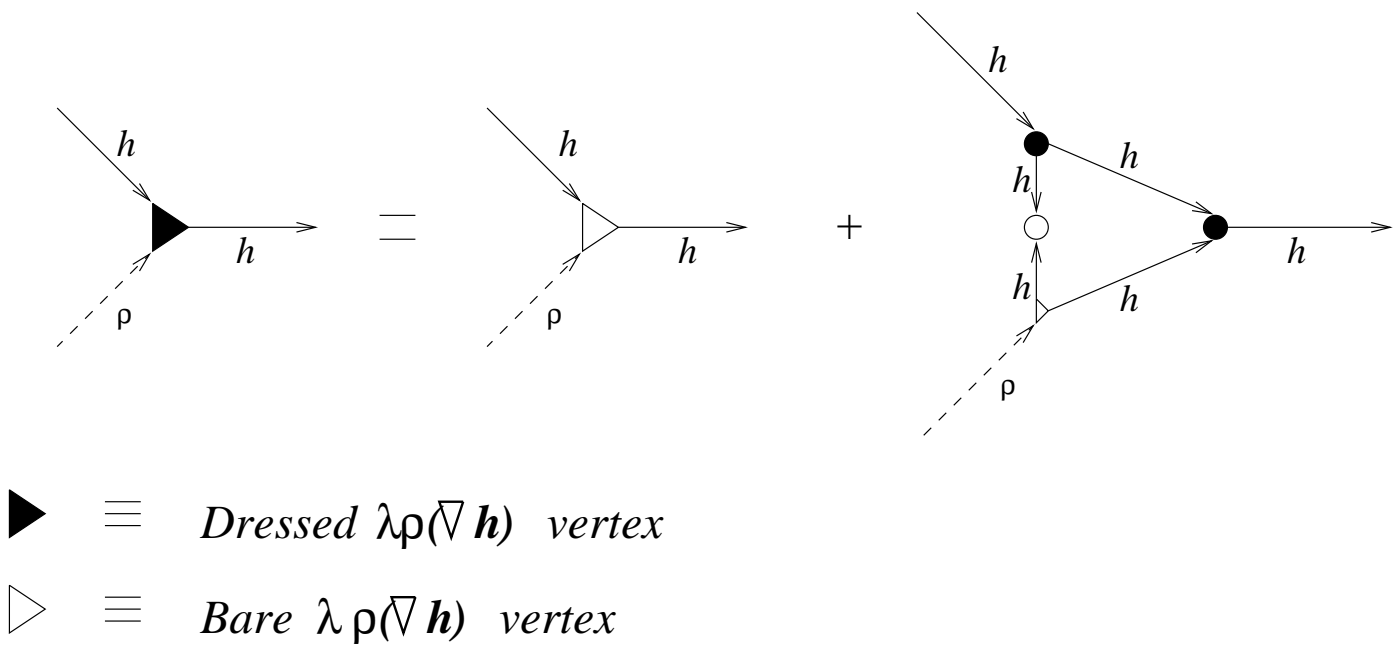
\begin{figure}
\caption{
One-loop  corrections to 
(a) the KPZ vertex, and
(b) the $\lambda$ vertex 
for the coupled equations of Case C (Eqs.(21)).
}
\label{fig-lambda_C}
\end{figure}

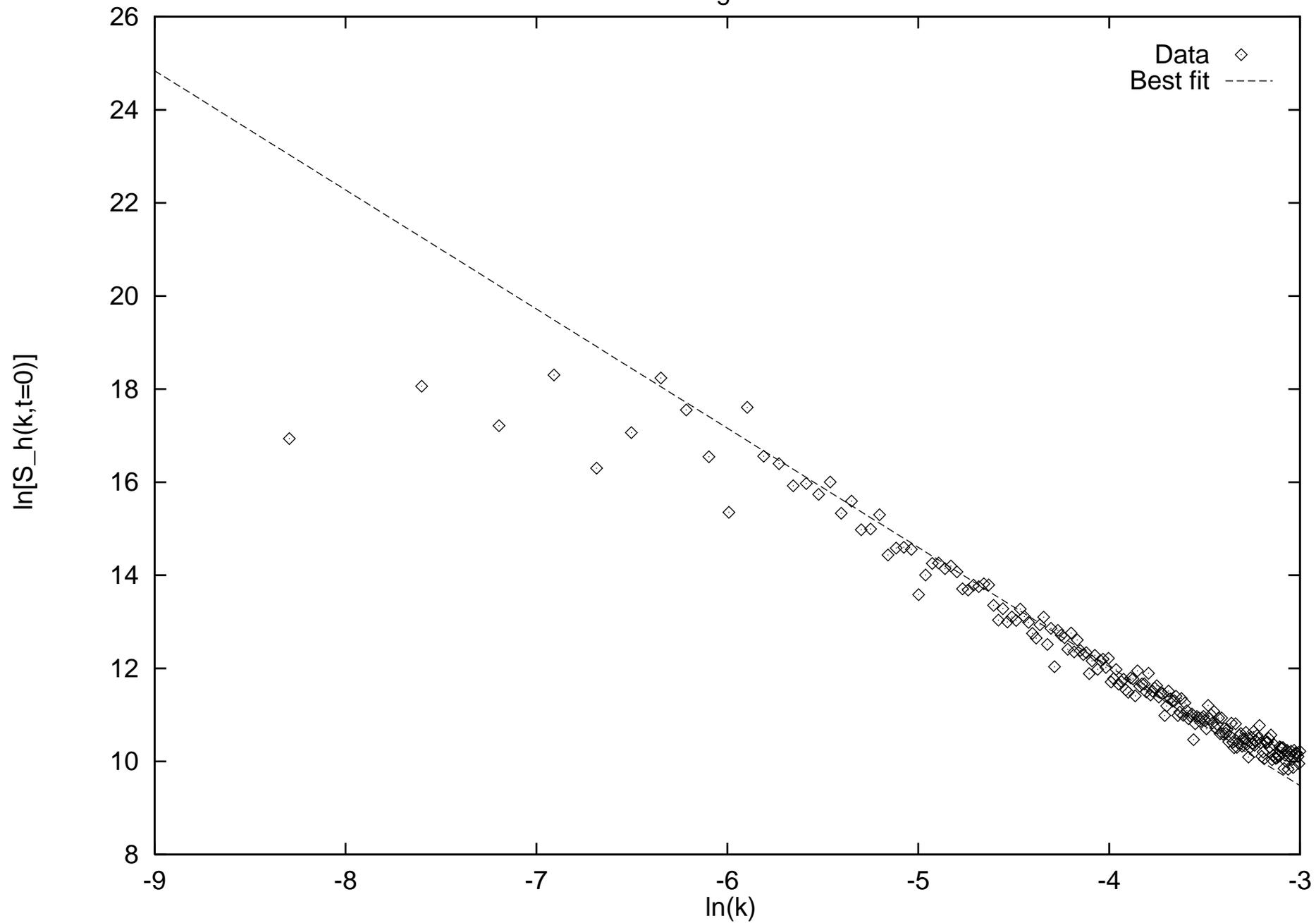
\begin{figure}
\caption{
Log-log plot of the single Fourier transform $S_h(k,t=0)$ vs $ k$  for  
Case C obtained from Eqs.(21). 
The slope of the fitted line is given by $-1-2\alpha_h = -2.56\pm 0.060$ 
The parameters used in the simulation are  $\nu = 10, \lambda = 1.0,
D_h = D_{\rho} = 1.0, \Delta_h^2 = 1.0$.
} 
\label{fig-C1}
\end{figure}

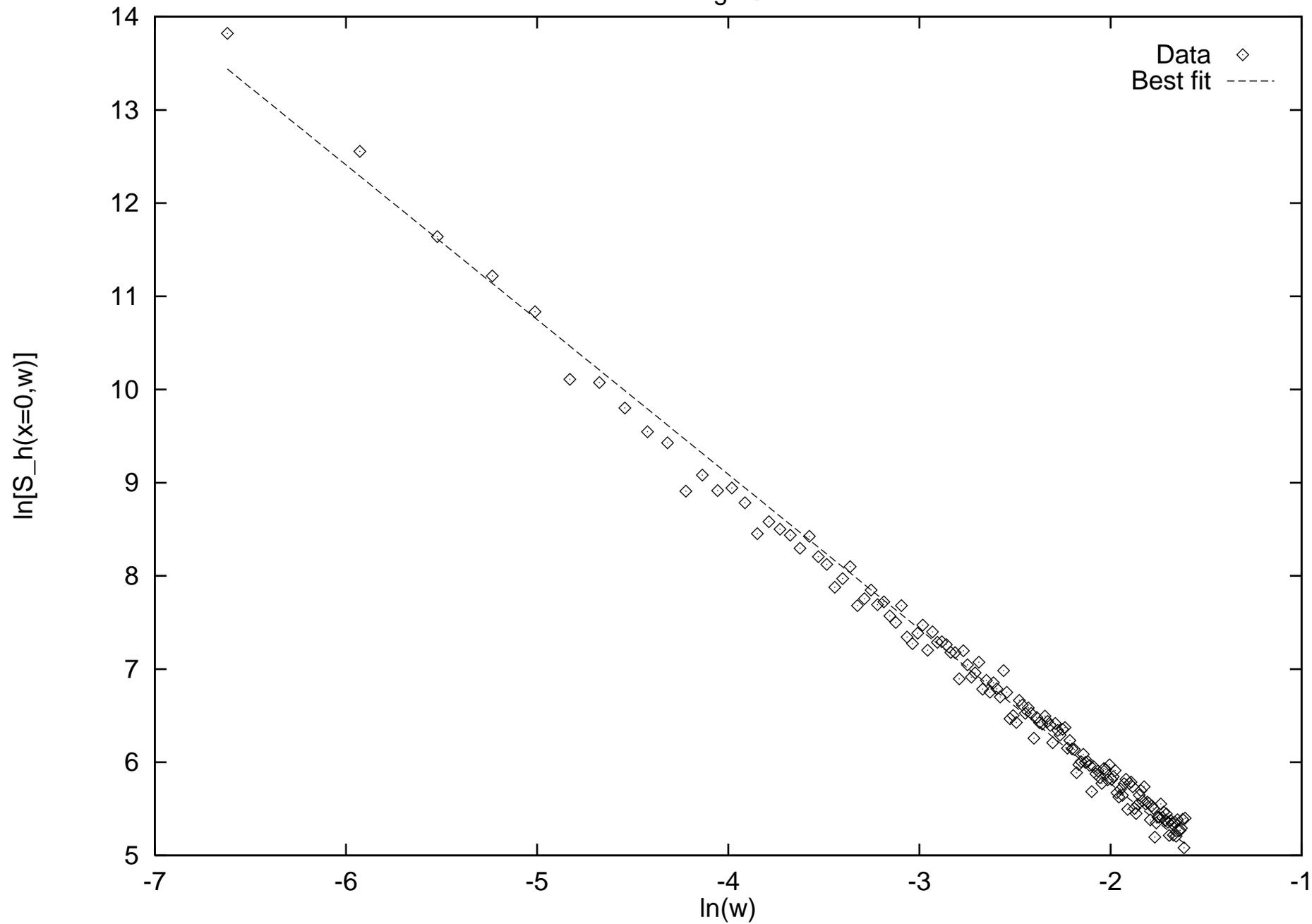
\begin{figure}
\caption{
Log-log plot of the single Fourier transform 
$S_h(x=0,\omega)$ vs $\omega$ obtained from Eqs.(21) ( Case C ). 
The best fit has a slope of
$-1-2\beta_h = -1.68\pm 0.011$ with parameters $\nu = 10, \lambda = 1.0,
D_h = D_{\rho} = 1.0, \Delta_h^2 = 1.0$.
}
\label{fig-C2}
\end{figure}

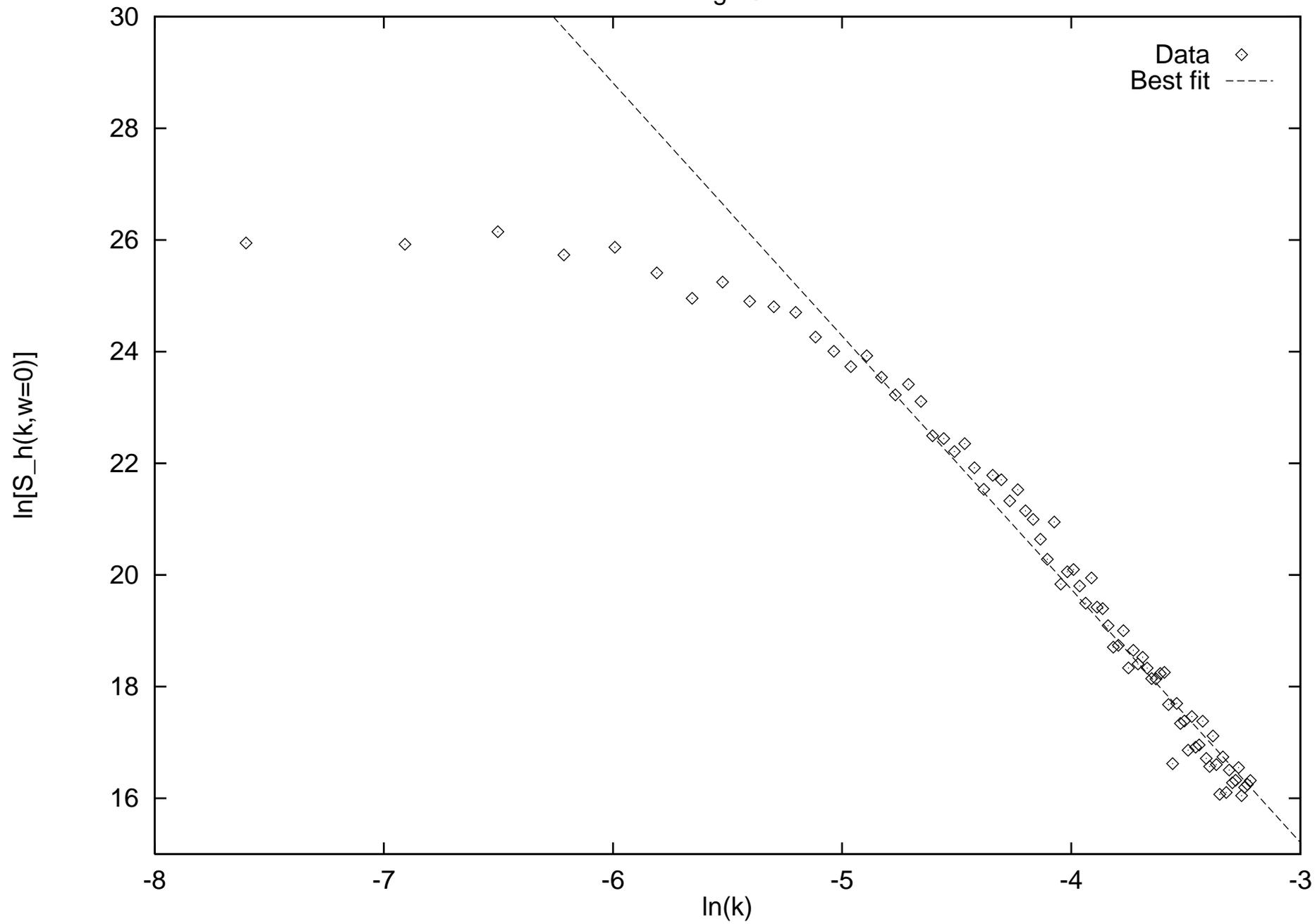
\begin{figure}
\caption{
Log-log plot of the double Fourier transform $S_h(k,\omega=0)$ vs $k$
obtained from Eqs.(21) ( Case C ). 
The best fit for high wavevector has a 
slope of $-(1+2\alpha_h+z_h) = -4.54\pm 0.081$. 
As $k \rarrow 0 $ we observe a 
crossover to slope of zero. 
Other parameters are $D_h = D_{\rho} = 1.0, \Delta_h^2 = 1.0, \nu = 10 $
and $\lambda = 1.0$.
}
\label{fig-C3}
\end{figure}

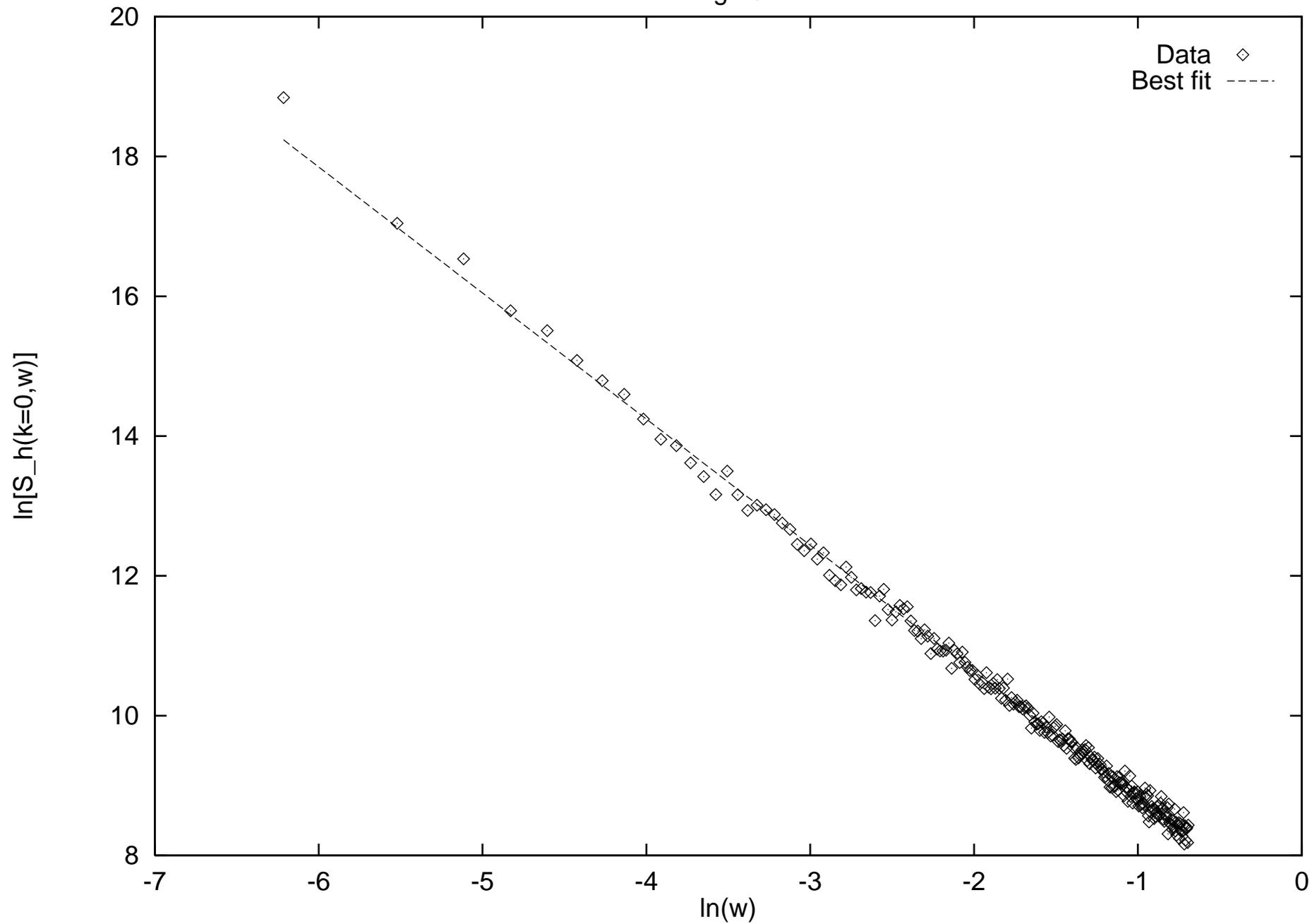
\begin{figure}
\caption{
Log-log plot of the double Fourier transform 
$S_h(k=0,\omega)$ vs $\omega$ obtained from Eqs.(21) ( Case C ).
 The best fitted line shown in the figure has a slope of  
 $-(1+2\beta_h+\frac{1}{z_h}) = -1.80\pm 0.007$. 
Other parameters are $D_h = D_{\rho} = 1.0, \Delta_h^2 = 1.0, \nu = 10$
and $\lambda = 1.0$.
}
\label{fig-C4}
\end{figure}

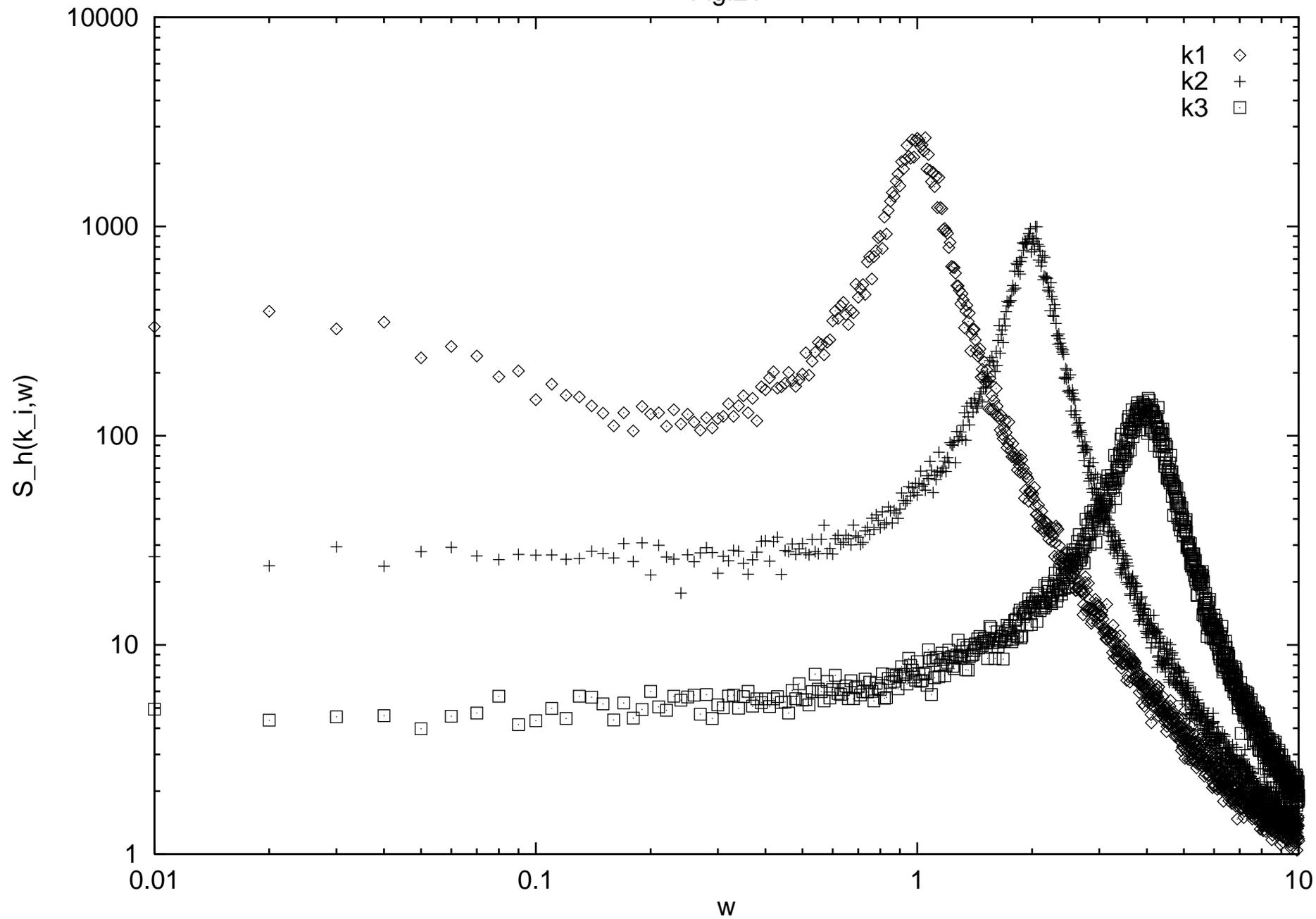
\begin{figure}
\caption{
The double Fourier transform $S_h(k_i,\omega)$ vs $\omega$ obtained 
from Eqs.(21) ( Case C ) evaluated at three 
different wavevectors $k_1 = 0.2 (\Diamond)$,
$k_2 = 0.4 ( + )$ and $k_3 = 0.8 (\Box)$ with parameters $D_h =
 D_{\rho} = 1.0, \Delta_h^2 = 1.0, \nu = 5
$ and $\lambda = 1.0$. The peaks correspond to frequencies 
$\omega_1 = 1.0$, $\omega_2 = 2.0,\omega = 4.0$.
}
\label{fig-C5}
\end{figure}

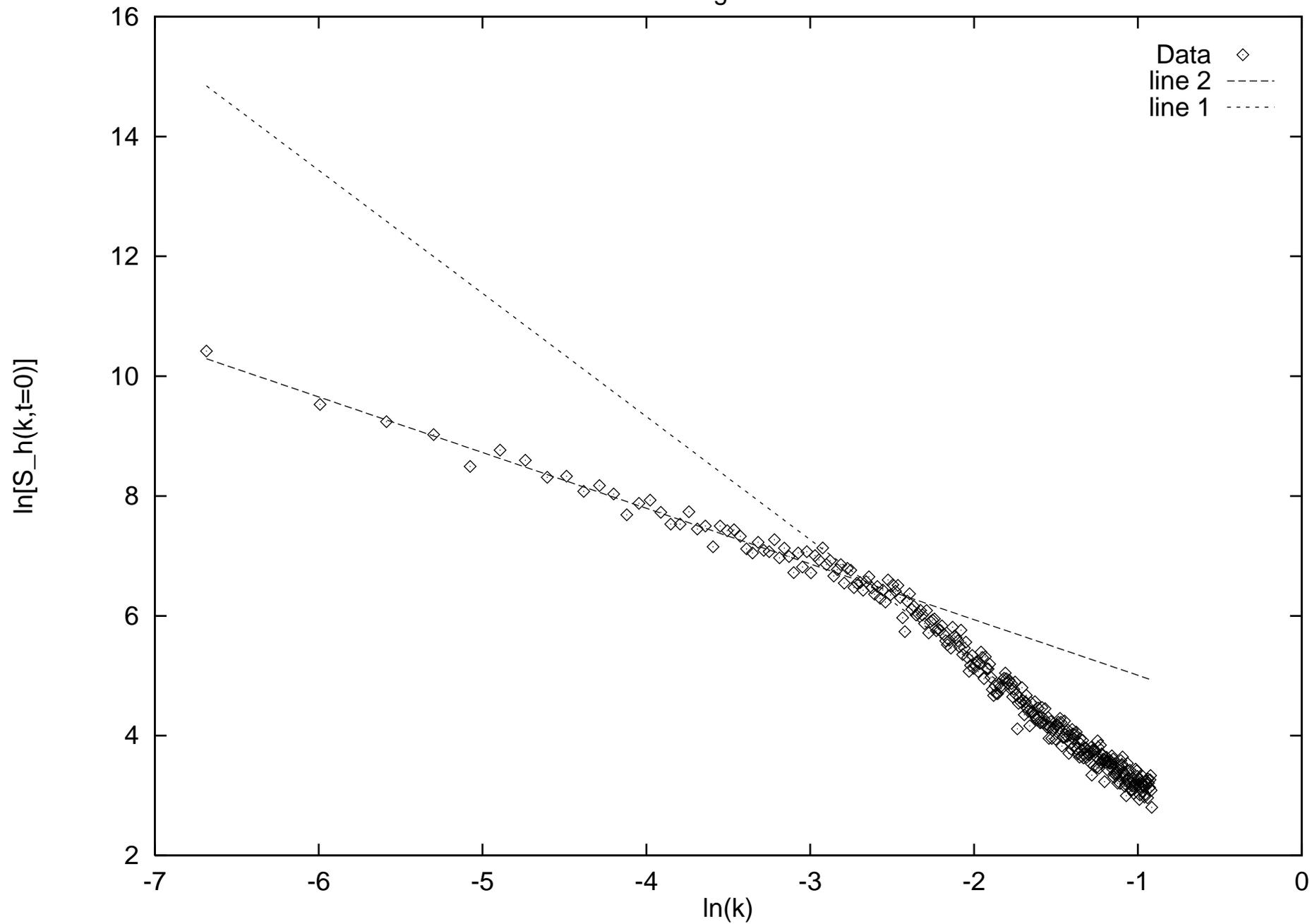
\begin{figure}
\caption{
Log-log plot of the single Fourier transform $S_h(k,t=0)$ vs $k$ 
obtained from the mean-field Eqs.(\ref{eq-bcre1},\ref{eq-bcre2}). The high $k$ region 
is fitted with a line of slope 
$-1-2\alpha_h = -2.05\pm 0.017$. The low $k$ region
is fitted with a line of slope $-1-2\alpha_h = -0.93\pm 0.024$. Note the  
crossover from $\alpha_h = 0.5$ at large $k$ to zero at small $k$. Other parameters are
$\nu^{\prime}=10$, $\lambda^{\prime} = 2.0$, 
$D_h = D_{\rho} = 1.0, \Delta_h^2 = 0.1$.}
\label{fig-bcre1}
\end{figure}

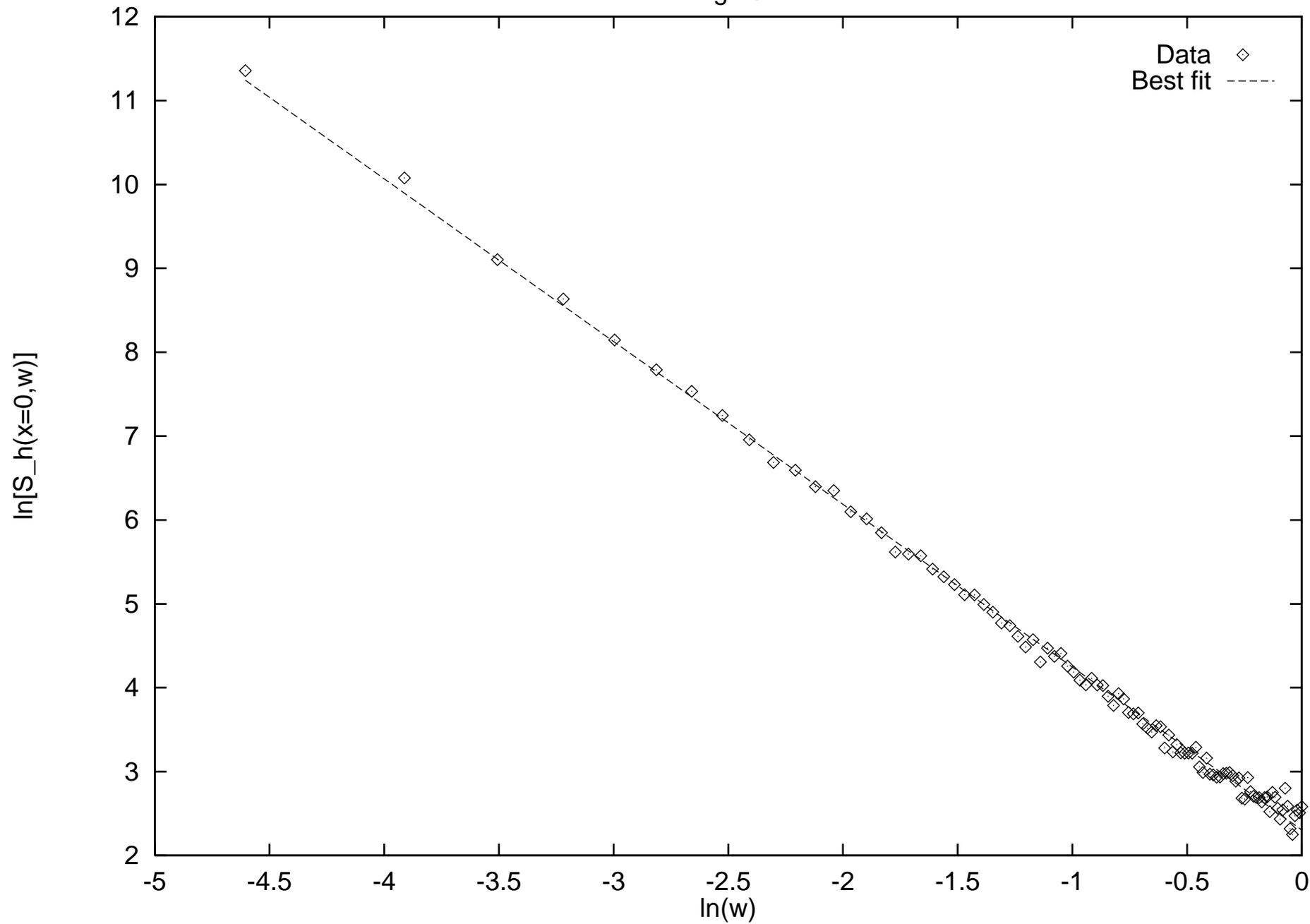
\begin{figure}
\caption{
Log-log plot of the single Fourier transform 
$S_h(x=0,\omega)$ vs $\omega$ for the mean-field Eqs.(\ref{eq-bcre1},\ref{eq-bcre2}). 
The best fit has a slope of 
$-1-2\beta_h = -1.94\pm 0.001$ with parameters
$\nu^{\prime}=10, \lambda^{\prime} = 2.0$,
$D_h = D_{\rho} = 1.0, \Delta_h^2 = 0.1$.
}
\label{fig-bcre2}
\end{figure}
\vskip 1.5cm
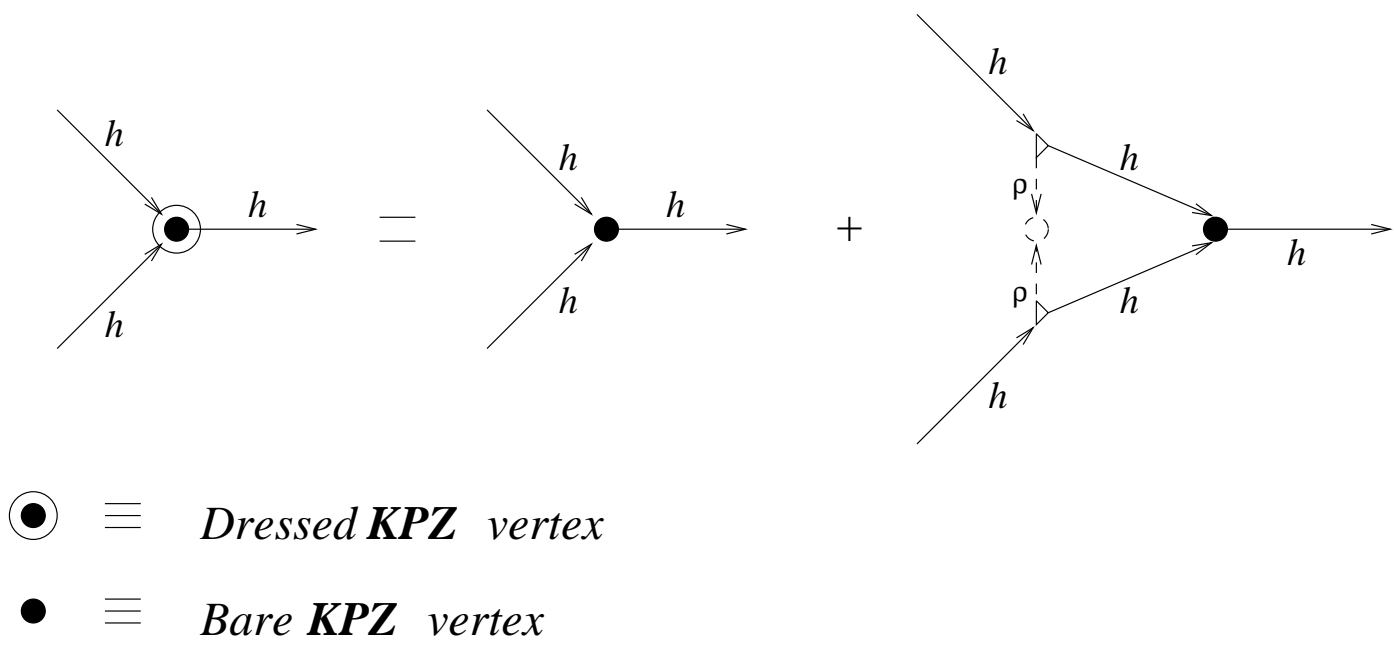
\end{document}